\documentstyle[12pt,aasms4]{article}

\lefthead{Ikebe et al.}
\righthead{The Structure of the X-ray Emitting Gas
in the Hydra-A Cluster of Galaxies}

\begin{document}

\title{THE STRUCTURE OF THE X-RAY EMITTING GAS IN THE HYDRA-A CLUSTER
OF GALAXIES}

\author{
Y. Ikebe\altaffilmark{1}, 
K. Makishima\altaffilmark{2},
H. Ezawa\altaffilmark{2}, 
Y. Fukazawa\altaffilmark{2}, 
M. Hirayama\altaffilmark{2,3}, 
H. Honda\altaffilmark{3}, \\
Y. Ishisaki\altaffilmark{2,4}, 
K. Kikuchi\altaffilmark{4},
H. Kubo\altaffilmark{2,3}, 
T. Murakami\altaffilmark{3}, 
T. Ohashi\altaffilmark{4}, 
T. Takahashi\altaffilmark{3}, \\
and K. Yamashita\altaffilmark{5}}
\affil{
ikebe@postman.riken.go.jp: maxima@miranda.phys.s.u-tokyo.ac.jp:
ezawa@tkyosf1.phys.s.u-tokyo.ac.jp: fukazawa@miranda.phys.s.u-tokyo.ac.jp:
hirayama@tkyosf1.phys.s.u-tokyo.ac.jp: honda@astro.isas.ac.jp:
ishisaki@phys.metro-u.ac.jp: kikuchi@phys.metro-u.ac.jp:
hkubo@miranda.phys.s.u-tokyo.ac.jp: murakami@astro.isas.ac.jp:
ohashi@phys.metro-u.ac.jp: takahasi@tkynext.astro.isas.ac.jp:
yamasita@satio.phys.nagoya-u.ac.jp}

\altaffiltext{1}{Cosmic Radiation Laboratory, The Institute of Physical and
Chemical Research (RIKEN), 2-1 Hirosawa, Wako-shi, Saitama 351-01, JAPAN}
\altaffiltext{2}{Department of Physics, University of Tokyo, 3-1, 
Hongo 7-chome, Bunkyo-ku, Tokyo 113, JAPAN}
\altaffiltext{3}{The Institute of Space and Astronautical Science, 
1-1, Yoshinodai 3-chome, Sagamihara-shi, Kanagawa 229, JAPAN}
\altaffiltext{4}{Department of Physics, Tokyo Metropolitan University, 
1-1, Minamioosawa, Hachioji-shi, Tokyo 192-03, JAPAN}
\altaffiltext{5}{Department of Physics, Nagoya University, 
Furoucho, Chikusa-ku, Nagoya-shi, Aichi 464-01, JAPAN}

\setlength{\baselineskip}{1.4pc}

\begin{abstract}
The temperature and abundance structure in the intracluster medium 
(ICM) of the Hydra-A cluster of galaxies is studied 
with {\it ASCA} and {\it ROSAT}.
The effect of the large extended outskirts in the point-spread
function of the X-Ray Telescope on {\it ASCA}
is included in this analysis.
In the X-ray brightness profile, 
the strong central excess above a single $\beta$-model,
identified in the {\it Einstein} and {\it ROSAT} data,
is also found in the harder energy band ($>$4~keV).
A simultaneous fit of five annular spectra
taken with the GIS instrument shows 
a radial distribution of the temperature and metal abundance.
A significant central enhancement in the abundance distribution
is found,
while the temperature profile suggests that the ICM is approximately
isothermal with the temperature of $\sim$3.5~keV.
The {\it ROSAT} PSPC spectrum in the central $1'.5$ region 
indicates a significantly lower temperature than the GIS result.
A joint analysis of the GIS and PSPC data reveals that 
the spectra can be described by a two temperature model as well as 
by a cooling flow model.
In both cases,
the hot phase gas with the temperature of $\sim$3.5~keV occupies
more than 90\% of the total emission measure within 
$1'.5$ from the cluster center.
The estimated mass of the cooler (0.5--0.7~keV) component 
is $\sim$2--6$\times 10^9$ M$_{\odot}$, which is comparable to 
the mass of hot halos seen in non-cD ellipticals.
The cooling flow model gives the mass deposition rate of $60 \pm 30$
M$_{\odot}$ yr$^{-1}$, an order of magnitude lower than the previous
estimation.
\end{abstract}


\section{INTRODUCTION}
X-ray imaging studies with the {\it Einstein} observatory and
the {\it ROSAT}
satellite revealed that many clusters exhibit central concentrations
in their X-ray brightness profiles.
The central excess emission is often interpreted as a result of 
a cooling flow, a thermal instability occurring at the densest part
of cluster (see \cite{Fabian94} for a review).
Edge et al. (1992) reported that 
about two thirds of all known clusters have such structure, 
suggesting that this structure is a fairly common characteristic
among clusters. 
Spectroscopic studies with {\it Einstein} and {\it ROSAT} found that
there are cooler gas components with temperature $\sim$10$^7$~K 
in cluster centers (e.g. \cite{Cani79}).
However, the previous spectroscopy is limited to the soft X-ray band 
($\sim$0.1--4~keV).
{\it ASCA} has provided the first opportunity to resolve X-ray spectra 
spatially in the wider 0.5--10~keV energy range.

In this paper we report the {\it ASCA} observation of the Hydra-A cluster
of galaxies ($z=0.0522$).
Its X-ray luminosity measured with the {\it Einstein} Observatory was 
4.1$\times$10$^{44}$ ergs~s$^{-1}$ (0.5--4.5~keV) (\cite{David90}), and
a consistent value of 4.8$\times$10$^{44}$ ergs~s$^{-1}$ in the same
energy band was indicated from the {\it Ginga} observation 
which measured the X-ray spectrum in the 2--10~keV band (\cite{Tsuru}).
These luminosities are among the largest of poor clusters 
(\cite{Kriss83}; \cite{Tsuru}). 
The average X-ray temperature of $\sim$4~keV, as measured with {\it Ginga}, 
is also rather high for a poor cluster.

The central region of this cluster is of particular interest. 
The cD galaxy 3C218 = Hydra-A is a strong radio source,
with a very complex radio morphology and the
highest Faraday rotation ever measured from a radio galaxy
(\cite{Ekers}; \cite{Kato}; \cite{Taylor}). 
Furthermore, the X-ray surface brightness profile obtained with the 
{\it Einstein} IPC exhibits a large central excess deviating 
from an isothermal $\beta$-model,
suggesting the presence of a cooling flow with an estimated mass 
deposition rate of 600 $\pm$ 120 $M_{\odot}$ yr$^{-1}$ (\cite{David90}).
This is one of the largest values thus far attributed to the X-ray emitting
gas in clusters of galaxies (\cite{Edge92}).

When we analyze the {\it ASCA} data, 
the complex response of the X-Ray Telescope (XRT; \cite{XRT}) 
onboard {\it ASCA}
introduces considerable difficulty for spatially resolved spectroscopy.
To confront with these difficulties, some analysis methods have been 
proposed (\cite{IkebeD}; \cite{MEA96}; \cite{Churazov96}).
The method employed in this paper is based on the similar idea in Markevitch
et al (1996) but is performed with different implementation.
Since the position resolution of {\it ASCA} is relatively poor,
the sharp {\it ROSAT} image is also very helpful 
to analyze the {\it ASCA} data.
Assisted by the {\it ROSAT} image, 
we measured the spatial distribution of the temperature and metal abundance
in the Hydra-A cluster from the {\it ASCA} observation.
In section 2, we describe the {\it ASCA} observation 
and the data selection procedure.
Section 3 gives the results using a conventional analysis method 
on the GIS and SIS data.
Section 4 describes the analysis method developed here.
In section 5, we analyze the X-ray brightness profiles in different 
energy bands.
In section 6, we describe the temperature and abundance structure studied 
using the GIS spectra,
and we also discuss the joint analysis of 
the {\it ROSAT} and {\it ASCA} data to study 
the temperature structure in the central region.
We discuss the results and summarize them in section 7.
$H_0 = 50\ {\rm km\ s^{-1}\ Mpc^{-1}}$ is assumed throughout.

\section{ASCA OBSERVATION AND DATA SELECTION}
{\it ASCA} (\cite{ASCA}) observed the Hydra-A cluster during the performance
verification phase on 1993 May 28 and 29.
The two GIS detectors (\cite{GIS-Ohashi}; \cite{GIS-Max}),
GIS-S2 and GIS-S3, were operated in PH normal mode. 
The two SIS sensors, SIS-S0 and SIS-S1, were operated in 4CCD faint 
or bright mode, 
and central region of the Hydra-A cluster was detected in the middle of 
chip 1 and chip 2 of SIS-S0, and chip 3 and chip 0 of SIS-S1.

We have applied the standard data cleaning procedure 
(see \cite{ABC})
to the GIS and SIS data, and selected the data with the criteria
that the cut off rigidity and elevation angle must be greater 
than 8 GeV/c and 5.0 degrees, respectively.
The additional selection criterion 
that the elevation angle from the sunlit earth is greater than 20 degree
was applied to the SIS data only.
In the present analysis,
we used 24.4 ksec of the SIS-S0 data, 22.7 ksec of the SIS-S1 data 
and 26.3 ksec of the GIS data survived the selection.
The X-ray intensity contour map from the summed GIS-S2 and S3 data
is shown in Fig.~\ref{gis-image}.
As the background
(cosmic X-ray background plus non X-ray background)
of the GIS data,
we utilized the data from the {\it Lynx field},
which was observed on 13--15 May 1993 when the spread discriminator 
in the onboard CPU of GIS was disabled as it was for 
the Hydra-A cluster (see \cite{GIS-Max}).
For the SIS background, we used the data of several blank sky fields 
observed during the PV-phase.

\section{ANNULAR SPECTRAL ANALYSIS}
For detailed study of the spatial variations in the spectra,
energy spectra are often accumulated from different regions
and are analyzed individually
(see e.g. \cite{Ohashi94}; \cite{Ikebe94}).
However, this conventional analysis method is no longer usable for
the {\it ASCA} data of clusters of galaxies.
Although the {\it ASCA} XRT has a sharply peaked Point-Spread Function
(PSF) with about 3 arcmin half-power diameter (HPD),
the PSF has wide outskirts with significant flux extending over 
the whole GIS field of view of 52 arcmin diameter (\cite{XRT}).
Therefore individual regions on the focal plane have inevitable 
contribution of the flux from other regions.
Moreover, such a flux-mixing effect depends considerably 
on the X-ray energy
because the PSF has wider outskirts at higher energies.
In particular, this energy-dependent flux-mixing effect 
is quite serious for the  study of clusters of galaxies.
Since their X-ray surface brightness decreases from center to 
periphery by several orders of magnitude,
the outer regions of clusters inevitably have a considerable
and energy-dependent flux from the inner, much brighter parts.
Therefore the meaning of the individual spectra accumulated over 
specified areas in the focal plane is very unclear.
Figure~\ref{mutual-contrib} demonstrates the mutual contribution 
of X-ray flux from annular sky regions 
to the annular regions on the focal plane
in case of the Hydra-A cluster.

Although not accurate, 
in this section we first perform model fittings
to individual annular spectra,
as have been done in previous papers on the {\it ASCA} data.
This is to illustrate the results when the energy-dependent
flux-mixing effect is not considered,
and to compare them with the results from the new analysis method
described in the subsequent sections.

The GIS and SIS spectra have been accumulated in concentric annular
regions centered on the cD galaxy, 
covering radius ranges of $0-1'.5$ (arcmin), $1'.5-3'$, $3'-5'$,
$5'-10'$, and $10'-20'$.
The data of GIS-S2 and GIS-S3 were then combined after the correction 
for the gain difference.
For the SIS, data from four different chips in the same detector were 
combined after relative gain corrections.
The SIS and GIS spectra thus obtained both exhibit strong emission lines 
from highly ionized ions.
We fitted annular spectra individually with 
a thermal bremsstrahlung model plus Gaussians modified by photo-electric
absorption, using the XSPEC package (ver. 9.0).
The energy ranges used for the fitting are 0.7--10~keV and 0.35--10~keV 
for the GIS and SIS, respectively.
For all the spectral fittings, we used the XRT effective area 
which is calculated assuming that the point source is being observed with
a circular detector of $6'$ radius size.
Figure~\ref{central-spec} shows the central $0-1'.5$ spectra 
together with the best-fit model.
At least three Gaussians 
at the rest-frame energies of 6.7~keV, $\sim$2.0~keV and $\sim$1.0~keV
are required,
which may correspond to the He-like Fe-K$_{\alpha}$,
He-and H-like Si-K$_{\alpha}$,
and a mixture of He-like Fe-L$_{\alpha}$ and H-like Ne-K$_{\alpha}$
emission lines, respectively.
There is also some indication of a line structure at $\sim$1.5~keV,
which may consist of He-like Fe-L$_{\beta}$ and H-like Mg-K$_{\alpha}$
emission lines.
The continuum is well described by a single temperature model.
When we performed the fitting using only below 3~keV energy band
or above 3~keV energy band,
the best-fit temperatures were not different.
The two temperature model did not improve the best-fit $\chi^2$
values at all.
We illustrate the projected radial profiles of the temperature 
and the equivalent width (EW) of the Fe-K emission lines 
in Fig.~\ref{brems+gauss_rpro}.
Unless otherwise specified, all errors are 90\% confidence for
one parameter of interest ($\Delta \chi^2 = 2.71$).
The two instruments have given consistent results, and
the gradual increase towards the outside is seen 
in the temperature profile.

We also fitted the spectra with the single-temperature
thin-thermal plasma emission model by Raymond and Smith 
(1977; hereafter RS-model).
The temperature, heavy element abundances, 
and normalization were varied as free parameters, 
while relative abundance ratios were fixed at the solar values given 
by Anders \& Grevesse (1989).
The projected radial profiles of temperature and abundance derived 
are shown in Fig.~\ref{wagiri-fit}.
The temperature profile obtained exhibits a gradual increase 
towards the outside and 
is consistent with the result from the fits with a thermal
bremsstrahlung model plus Gaussians (Fig.~\ref{brems+gauss_rpro}-(a)).
The abundance values obtained with the SIS are significantly
smaller than those obtained with the GIS (Fig.~\ref{wagiri-fit}-(b)).
The same systematic difference has been found in the data 
from various clusters observed with ASCA 
(\cite{Fukazawa96}).
This may be because the actual SIS energy resolution is slightly
worse than the prediction of the response matrix, probably because 
of the chip summation.
When we applied an alternative plasma emission model,
Mewe-Kaastra model (MEKA-model: \cite{MEKA1}; \cite{MEKA2}; \cite{MEKA3}),
we derived consistent results with those from the RS-model.
Consistency between RS-model and MEKA-model results has been found 
from the ASCA data of other clusters with temperature of 
$\sim$3--4~keV (\cite{Mushotzky96}).
As already pointed out, we cannot interpret the radial profiles
presented in Fig.~\ref{brems+gauss_rpro} and Fig.~\ref{wagiri-fit}
as true projected structure on the sky.
The gradual increase towards the outside seen in the temperature profiles
is not a real temperature structure but is likely to be an artifact 
due to the XRT responses.
Simulations assuming an isothermal distribution also produce
a similar temperature profile (\cite{Takahashi95}).

\section{ANALYSIS METHOD FOR THE EXTENDED SOURCES}
\subsection{Analysis Methods}
One way to analyze extended sources
is to perform simultaneous fitting of the X-ray image and 
the spectra;
that is, 
simultaneous fitting of the spectra extracted from different regions, 
or similarly, simultaneous fitting of the images sorted in different 
energy bands.
We start with an X-ray source model which, in case of a cluster,
describes spatial distributions of the X-ray emissivity, 
temperature, and abundance.
A given model is converted to the simulated {\it ASCA} data through
the convolution calculations with the instrument responses
which include the XRT PSF and effective area,
as well as the quantum efficiency
and energy redistribution function of the GIS or SIS.
The goodness of fit of the model is evaluated by comparing 
the simulated data with the observed data, and
the results are iteratively fed back to the input model.

The XRT+GIS PSFs used in our analysis were produced by interpolating 
the actually obtained X-ray images of Cyg~X-1 with GIS.
However, the Cyg~X-1 is so bright that 
the SIS can not measure its brightness profile precisely 
due to event pile-up;
thus can not reproduce the XRT+SIS PSFs with the same method.
Therefore, in this paper we analyzed only the GIS data 
with the analysis method described here.
The detailed description of PSF calibration is in the Appendix.

A technical problem in this analysis is how to reduce the calculation
time.
The image convolution, which is the most time consuming part,
is mathematically expressed as an operation by the 2-dimensional
matrix called {\it the image response matrix}
with size as large as $\sim 20000 \times 20000$
for the GIS full-resolution image.
To reduce the calculation time, 
we have developed an analysis method using the Monte-Carlo technique
(\cite{Takahashi95}; \cite{IkebeD})
which is often much faster than the huge matrix operation.
We only need to generate events more numerous, say by a factor of 10,
than the actual events.
This Monte-Carlo method was employed for the analysis of 
the Fornax cluster (\cite{Ikebe96}),
the Centaurus cluster (\cite{IkebeD}),
3A0336+098, A1795, MKW3s and PKS2354-35 
(\cite{Ohashi95}; \cite{Kikuchi95}),
Ophiucus cluster (\cite{Matsuzawa}), and Coma cluster (\cite{Honda}).

\subsection{Image Response Matrices}

An alternative method to reduce the calculation time is to reduce
the size of the image response matrix.
Due to the limited number of observed photons,
we usually bin the image so that 
each image pixel has sufficient number of events 
that $\chi^2$ statistics can be employed.
Moreover, an X-ray source model given is often so simple 
that the number of sky regions can be largely reduced.
Thus, the dimension of the image response matrix 
can be reduced drastically.
Since the Hydra-A cluster can be reasonably modeled with a circularly 
symmetrical model,
as will be described in the following sections,
we employed the matrix operation to generate simulated data
rather than using the Monte-Carlo method.
This method was first proposed by Markevitch et al (1996) and
had been applied to the {\it ASCA} data analysis of
A2163, A754, A2256, A2319, A665, Triangulum Australis, A3558, and AWM7
(\cite{MEA96}; \cite{HM96}; \cite{Markevitch96}; \cite{MSI96}; \cite{MV97}).
Here we recall the formula of the matrix operation and define the $\chi^2$
function for the minimum $\chi^2$ fitting.

The model prediction is given by the form:
\begin{equation}
M\!O\!D\!E\!L^{det}_{\ \ pi} = \sum_{energy} \sum_{sky} 
R\!M\!F_{pi,\ energy}\ I\!R\!M^{det,\ sky}_{\ \ energy}\ S^{sky}_{\ \ energy}\ ,
\end{equation}
where $I\!R\!M$ is the image response matrix, which is defined as:
\begin{equation}
I\!R\!M^{det,\ sky}_{\ \ energy} = P\!S\!F^{det,\ sky}_{\ \ energy}\ E\!F\!F^{sky}_{\ \ energy}\ .
\end{equation}
Here, $S$ is the initial cluster model on the sky
which gives the X-ray surface brightness as a function of X-ray energy
(expressed with the suffix $energy$) and sky position ($sky$);
$E\!F\!F$ is the effective area including quantum efficiency of the focal
plane instruments expressed as a function of X-ray energy 
and sky position;
$P\!S\!F$ is the XRT PSF which gives the photon distribution 
on the detector ($det$) depending on the X-ray energy and sky 
position;
and $R\!M\!F$ is the energy redistribution function which expresses 
pulse height ($pi$) distribution for an incident monochromatic-energy
X-ray event.
While both the {\it ASCA} data and the initial cluster model 
on the sky ($S$) have three dimensions, 
the fitting is performed in the form of some energy-sorted brightness
profiles, or in the form of some ring-sorted spectra.

To evaluate the best-fit parameters of the assumed model,
we defined the $\chi^2$ function as:
\begin{equation}
\chi^2 = \sum_{det} \sum_{pi} 
\frac{(D\!A\!T\!A^{det}_{\ \ pi} 
- B\!G\!D^{det}_{\ \ pi} - M\!O\!D\!E\!L^{det}_{\ \ pi})^2}
{(\sigma^{det}_{\ \ pi})^2 }
\end{equation}
where
\begin{equation}
(\sigma^{det}_{\ \ pi})^2 
= \sigma_{D\!A\!T\!A^{det}_{\ \ pi}}^2 + \sigma_{B\!G\!D^{det}_{\ \ pi}}^2 
+(a \times B\!G\!D^{det}_{\ \ pi})^2 +(b \times M\!O\!D\!E\!L^{det}_{\ \ pi})^2 \ .
\end{equation}
Here $D\!A\!T\!A$, $B\!G\!D$ and $M\!O\!D\!E\!L$ 
are the values of the observed data counts,
the normalized background, and the model, respectively.
$\sigma_{D\!A\!T\!A^{det}_{pi}}$ and $\sigma_{B\!G\!D^{det}_{pi}}$ 
are statistical errors associated with respective quantities, 
while the latter two terms in eq.~(4) represent systematic errors.
The systematic error of the image response matrix is included 
in the $\chi^2$ value by adding 5\% ($b = 0.05$) of the model flux.
We also introduced 10\% of the background intensity estimated from 
the blank sky data in the systematic errors; that is $a = 0.1$.

We have checked the consistency between the Monte-Carlo method
and results from the image response matrices.
For example, from a best fit model obtained by fitting using 
the image response matrices, the Monte-Carlo simulation can reproduce
the data which agree with the actual data with sufficient accuracy.

\section{RADIAL BRIGHTNESS PROFILE}
\subsection{ASCA Data and Fitting Results}
Using the analysis method described in \S~4.2, 
we first analyzed the X-ray images taken with the GIS.
The X-ray image has a good circular symmetry, and
the azimuthally-averaged background-subtracted
radial brightness profiles in different energy ranges 
are shown in Fig.~\ref{gis-rpro}.
They are centered on the peak of the X-ray intensity 
and have 20 bins in total of 1 arcmin width each.
With {\it ASCA}, we were able to obtain the image in 
the 4--10~keV range for the first time, 
and these radial profiles are very similar among different energy 
bands.

In order to perform the model fitting to the radial profile
following the method described in \S~4.2,
we constructed image response matrices of a size
26 $\times$ 20 for each of the 201 energy bins.
Each matrix represents flux contribution from 26 sky regions
to the 20 regions on the focal plane at a specified energy.
The 26 sky regions employed here are 8 annular regions of $0'.25$ 
width for $r<2'$
and 18 annular regions of $1'$ width for $r=2'-20'$,
while the detector region was divided into 20 annular bins 
of $1'$ width each.
Since the matrix used for the fitting represents contribution 
only from sky regions within $20'$ radius,
it is implicitly assumed that no emission is generated outside 
of the $20'$ radius.
This assumption is consistent with the {\it ROSAT} image 
which will be described in the next section, 
and even when we assume the maximum radius
to be larger than $20'$, 
the fitting results do not change significantly.

As the model brightness profile, we employed an empirical $\beta$ 
model expressed as:
\begin{equation}
\Sigma(r) = \Sigma_0 
	\left[ 1 + \left(\frac{r}{r_c}\right)^2 \right]^{-3\beta + 0.5}\ ,
\end{equation}
where $r$ is the projected angular distance from the center, 
$\beta$ represents the beta parameter
and $r_c$ is the core radius.
We assume that the energy spectrum takes the same form in the entire
cluster and is expressed with the RS-model modified by the Galactic 
absorption.
The temperature, heavy element abundance, and the hydrogen column
density are assumed to be 3.4~keV, 
0.5~solar and $6\times10^{20}$ cm$^{-2}$ respectively,
as derived in \S~3 from the spectral fitting of the GIS data 
extracted from the inner most region.
The free parameters are $r_c$, $\beta$ and $\Sigma_0$.

With this single $\beta$-model, 
we fitted the individual radial profiles in Fig.~\ref{gis-rpro}
and obtained acceptable fits for all energy bands.
The best fit parameters are summarized in Table~\ref{rprofit}.
We can compare this with the results from the {\it Einstein} IPC;
that is, the detection of the central excess emission above 
a $\beta$-model (\cite{David90}).
Since {\it ASCA} has a much poorer angular resolution than 
the {\it Einstein} IPC,
the central excess in the IPC data may be unresolved in the 
{\it ASCA} data.
Thus we obtain a single $\beta$-model with core radius smaller 
than $1'.6$,
consistent with the results obtained from IPC data in the $r>1'.5$
region (\cite{David90}).
In fact, 
the GIS radial profiles are well reproduced by convolving 
the {\it ROSAT} PSPC image with the {\it ASCA} response (see \S~5.3).
The important result here is that all the radial brightness profiles 
in different energy bands show an almost identical shape,
as can be seen from Table~\ref{rprofit}.
This suggests that 
the central excess emission may be present in higher energy bands 
as well as in the soft energy band.
We will come back to this issue later.

\subsection{ROSAT Data and Analysis}
As described in \S~4, to obtain the temperature structure 
in the Hydra-A cluster,
we need to fit X-ray image and spectra simultaneously.
However, the brightness profile obtained from 
the GIS data alone would be highly model dependent because 
of the relatively wide PSFs.
An alternative preferable way is to utilize the {\it ROSAT} 
image itself as the model of the brightness profile, 
as has been done in Markevitch et al. (1996).
We thus analyzed the data of the {\it ROSAT} PSPC.

The {\it ROSAT} PSPC observed the Hydra-A cluster 
from 1992 Nov 8 to 1992 Nov 9
with a total on-target exposure time of 18~ksec and the pointing 
position of $9^h18^m05^s, -12^{\circ}06'00''$ (J2000).
We obtained the processed data of the PSPC from the archival 
data base provided by the {\it ROSAT} Guest Observer Facility in NASA 
Goddard Space Flight Center.
For the data reduction we followed the procedure described 
by Snowden et al (1994) and used their software (\cite{Snowden95}).

From the PSPC data, 
the valid time is selected so that the Master Veto count rate 
is always smaller than 170 counts s$^{-1}$ to eliminate the data with
anomalously high particle background rates 
(\cite{Snowden92}; \cite{Plucinsky93}).
Using Snowden's software we can estimate the count rate of the four 
background components, the particle background (PB), 
after-pulse events (AP), scattered solar X-ray background (SB),
and long-term enhancements (LTE).
After subtracting the background, the PSPC image was corrected 
using the exposure map 
which represents the effective area and the exposure time as 
a function of the position in the sky coordinates.
We thus obtained the background-subtracted flat-field images 
in the 0.14--2.04~keV band (Fig.~\ref{pspc-image}).
In the PSPC data, X-ray flux from the cluster is detected out to
$\sim$20 arcmin.

The X-ray image taken by the PSPC exhibits very good circular 
symmetry.
To express the X-ray brightness profile quantitatively,
we performed model fitting to the radial brightness profiles.
We assumed that the PSPC point-source responses are ideally 
point-like;
this is true if we do not discuss structures smaller than $\sim$1$'$.
Firstly we fit the PSPC radial profile with a single-$\beta$ model,
but the fit was quite poor.
As a next step, we applied a sum of two $\beta$-models to 
fit the profile
and derived good fit as shown in Fig.~\ref{pspc-rpro-fit}.
The best fit parameters of the two $\beta$-models are
$(r_c, \beta)$ = $(1'.36, 2.5)$ for the narrower component, 
and $(r_c, \beta)$ = $(1'.80, 0.68)$ for the wider component.
In the narrower component the intensity is reduced by half 
at $r=0'.44$.
The normalization ratio of the two $\beta$-models,
$f \equiv$ (normalization of the narrow $\beta$-model)
         / (that of the wide $\beta$-model),
is 4.9.
This reconfirms the existence of the central excess emission found
with the {\it Einstein} IPC.

\subsection{Combined ASCA and ROSAT Data}

In the soft X-ray band where {\it ROSAT} and {\it ASCA} 
are both sensitive,
the {\it ASCA} radial profile also must be fitted with
the same double $\beta$-model which fit the PSPC radial profile.
Thus, we fit the GIS radial profile in 0.5--2.5~keV with the double 
$\beta$-model, in which the two sets of $\beta$ parameters 
and core radii
($r_c$), and the ratio between the two normalizations ($f$) were fixed
at the best fit values derived in \S~5.2.
We derived a good fit
and the obtained minimum chi-square value was 20.5/($\nu$=19).

In the higher energy bands, 
does the radial profile require a central excess?
In the previous section, we suggested that the similar radial 
profiles in
different energy band could be the evidence of the existence of 
the central excess emission in the higher energy bands as well.
Using the double $\beta$ model, we have checked if the higher energy
bands also show the central excess or not.
In this case, we let the normalizations for the two $\beta$-model 
components to vary independently, 
while the two sets of $\beta$ parameters and core radii
were fixed to the values which were obtained in \S~5.2.
The fitting results are illustrated in Fig.~\ref{norm-ratio}.
In all energy ranges, 
the normalization ratios ($f$) obtained are consistent with 4.9
which is the best fit value obtained with the PSPC radial profile 
in \S~5.2.
Moreover, when we fixed the normalization of 
the narrow $\beta$-component to be 0, the fit became unacceptable;
the minimum $\chi^2$ values are 101, 133, 81, 48, and 30,
with $\nu=19$
for the 0.5--1.5~keV, 1.5--2.5~keV, 2.5--4~keV, 4--6~keV, 
and 6--10~keV, respectively.
These results imply that the narrower of the two $\beta$-model 
components is definitely needed in individual radial profiles,
not only in the soft X-ray band but also in the harder energy ranges.
Therefore, we can conclude that the central excess emission 
is most likely to present in the higher energy band above 4~keV
as well as in the soft X-ray band.

\section{TEMPERATURE AND ABUNDANCE STRUCTURE}
\subsection{Temperature and Abundance Profile}
In the last section, we studied the radial brightness profiles 
in different energy ranges.
In this section, we describe the temperature and abundance profile 
based on the spatially sorted energy spectra.
Using the method described in \S~4,
we performed simultaneous fitting to the five GIS annular spectra
which were fitted individually in \S~3,
fully accounting for the contributions of the X-ray flux 
from the sky to the focal plane.
The image response matrices used for this analysis have the dimension
of 26 $\times$ 5 for each of the 201 energy bins.
Twenty-six model regions consist of 8 annuli each $0'.25$ wide 
for $r=0-2'$ and 18 annuli each $1'$ wide for $r=2'-20'$.

As the model brightness profile,
we employed the PSPC radial brightness profile obtained in \S~5.
We assumed that the PSPC image represents the surface brightness
profile in the energy range below 2~keV,
and converted those surface brightness to emission measures
according to the plasma emission code for a given temperature
and abundance.
The projected profiles of the temperature and abundance were assumed 
to be constant within the individual 5 annular regions 
of $0-1'.5$, $1'.5-3'$, $3'-5'$, $5'-10'$, and $10'-20'$,
which correspond to the same five annular regions used in 
extracting the GIS spectra in detector coordinates (Table~1).
The hydrogen column density is assumed to be constant over the entire 
cluster.
Therefore, the free parameters are 5 temperatures,
5 abundances,
the hydrogen column density
and the overall normalization.

The best fit parameters are summarized in Table~\ref{gis-doujifit}
and the temperature and abundance profiles are illustrated 
in Fig.~\ref{gis-doujifit-fig} by crosses.
Except for the central region, there seems to be an indication of
gradual temperature decrease towards the outside, 
but it is not statistically significant.
In the abundance distribution, only the central value was well 
constrained
while only upper limits were obtained for the outer regions.
However, the central abundance value is larger than all other best-fit
values in the outer regions.
This suggests there is a central concentration in the abundance
distribution.
In order to study the difference between the central $1'.5$ region and
the outside region more clearly, 
we again fitted the five GIS spectra simultaneously
assuming a common temperature and abundance outside $1'.5$.
The results are summarized in Table~\ref{gis-doujifit-2} and 
illustrated in Fig.~\ref{gis-doujifit-fig} with diamonds.
The central $1'.5$ abundance value is significantly larger than that
of the outside, implying there is a central concentration 
in the abundance distribution.

\subsection{Central Cool Region}
Since the radiative cooling time in the central region of the Hydra-A
cluster is shorter than the Hubble time ($2 \times 10^{10}$ yr),
the presence of a cooling flow 
is expected as discussed by David et al. (1990).
If there is no significant heat input, the thermal instability 
occurs in the central densest part of the cluster gas and 
a rapid radiative-cooling process would lead to 
the formation of a cooler gas phase surrounded by the hotter ICM.
As presented in the previous sections, 
the {\it ASCA} spectral analysis showed 
that the X-ray emission from the central $1'.5$ radius region
can be attributed to the 3.4~keV gas and 
there is no clear evidence of a central cooler gas component.
However, 
the GIS and SIS efficiencies decrease so rapidly below 1~keV
that {\it ASCA} is not very sensitive to emission 
from gas cooler than $\sim$1~keV.
The {\it ROSAT} PSPC is sensitive down to $\sim$0.1~keV
and thus these data would be useful to investigate 
whether there is an additional central cooler gas component.
In the following, we describe the results from the PSPC spectral 
fitting and joint analysis of the PSPC and GIS spectra.

We analyzed the {\it ROSAT} PSPC data first.
From the same data set selected in \S5.2, 
we accumulated the pulse-height spectrum in the central $1'.5$ region.
Each X-ray event was weighted using the vignetting function
so that the X-ray image became flat.
The original 256 channel bins are summed up into 7 energy bands named 
R1L through R7 defined in Snowden et al. (1994).
For the model fitting, 
we used only the 6 bins of R2 through R7 to avoid the after-pulse 
event contamination which could be contributing to the R1L band.
The background spectrum was made from the annular region of 
$r=36'-46'$ where no apparent point source was found.

We fitted the PSPC spectrum thus obtained with a single temperature
RS-model as well as the MEKA-model, 
with the response matrix named "pspcb\_gain2\_256.rsp" 
publicly available from NASA Goddard Space Flight Center.
The best-fit temperature, abundance, and hydrogen column density
and their 90\% confidence errors for one parameter of interest
obtained with the RS-model were 
2.27 (2.03--2.63)~keV, 0.30 (0.21--0.42)~solar,
and 3.79 (3.59--3.99) $\times 10^{20}$ cm$^{-2}$, respectively.
MEKA-model gave 2.25 (2.00--2.55)~keV, 
0.33 (0.23--0.45)~solar, and 3.78 (3.58--3.98)~cm$^{-2}$.
The derived temperatures are significantly lower than that obtained 
from the GIS data for the corresponding central region.
This temperature discrepancy between the GIS and the PSPC spectra 
indicates that the emission consists of multiple temperatures.
This is not surprising, because, even if the central cluster region
is filled with cool plasma, there must be an inevitable contribution
from the foreground and background regions of the cluster where 
the temperature is $\sim$3.4~keV.

As a next step, 
we performed simultaneous fitting of the PSPC and GIS spectra.
Because of the broad XRT PSF, 
the derived GIS spectrum within $1'.5$ radius consists of 
X-ray flux not only from the corresponding
sky region but also from outer regions of the sky.
However, the contamination from outside the $0-1'.5$ sky region
does not significantly affect the results.
This is clear from the fact 
that the annular spectral analysis performed in \S~3
and the simultaneously fitting of the five GIS spectra in \S~6.1
gave the same results for the central $1'.5$ region.
Therefore, 
we used only the central $0-1'.5$ GIS spectrum for the joint analysis
with the PSPC spectrum.
The effective area for the GIS spectrum was calculated from the PSPC 
image within $1'.5$ convolved with the GIS+XRT PSFs.
That effective area represents the contribution from the central
$1'.5$ sky region to the $1'.5$ radius detector region.

We found that the single temperature model
can not explain the PSPC and GIS spectrum simultaneously.
A single temperature RS-model and MEKA-model gave minimum $\chi^2$ 
values of 89.9 and 92.4, respectively, for 60 d.o.f.
There must be at least two different temperature components.
We attempted the following two types of models to fit the PSPC and
GIS spectra simultaneously.
Model 1 is a two temperature model, that is, two different temperature
components coexist within $1'.5$. 
The fitting parameters are the two temperatures, 
their emission measures, common abundances, and the galactic
absorption.
We allow the emission measures of the hot component for the PSPC
and GIS spectra to vary independently,
but constrain all other parameters to be equal.
We employed the RS-model and MEKA-model as the plasma emission code.
We derived acceptable fits in both cases.
The best fit model is shown in Fig~\ref{2temp-model-fig}, 
and the derived parameters are summarized in Table~\ref{2temp-model}.
The fraction of the cool component, 
the cool component percentage in the emission measure (EM),
defined as $F_{cool} \equiv EM(T_{cool})/EM(r<1'.5)$,
was $\sim$2\%.
Fig.~\ref{f-tcool}-(a) shows the $\chi^2$ contours 
in the $F_{cool}-T_{cool}$ space.

Model 2 is a projected cooling flow model 
which is expressed as $CF(\dot{M},T_{max},T_{min}) + P(T_{max})$,
where $CF$ is the cooling flow model 
by Mushotzky \& Szymkowiak (1988),
and $P$ represents the additional isothermal component.
$CF$ is a function of the mass deposition rate $\dot{M}$, 
the maximum temperature $T_{max}$ from which the gas cools,
and the minimum temperature $T_{min}$ to which the gas cools.
The temperature is distributed continuously from $T_{max}$ 
to $T_{min}$
and each temperature component has emission measure which is
inversely proportional to the total emissivity of that temperature.
The additional isothermal component must have the same temperature 
as $T_{max}$. 
In this model 2, we also applied two different emission codes,
RS-model and MEKA-model 
(CFLOW and MKCFLOW XSPEC model, respectively, for $CF$ component)
and derived acceptable fits (Table~\ref{cflow}).
The chi-square contour map for $\dot{M}$ vs $T_{max}$
is shown in Fig~\ref{f-tcool}-(b).
The estimated mass deposition rate is 60$\pm$30 M$_{\odot}$ yr$^{-1}$,
while the fraction of the cooling flow component to the total emission
measure within the $1'.5$ region ($F_{CF} \equiv EM(CF) / EM(r<1'.5)$)
was $\sim$8~\%.
With both model 1 and 2, 
the obtained X-ray luminosity within $1'.5$ region was 
$2.6 \times 10^{44}$ ergs s$^{-1}$ in the 0.5--4.5~keV band, 
and is primarily the emission of the hot non-cooled component with 
the temperature of $\sim$3.5~keV.
The 0.5--4.5~keV luminosity of the cool component of model~1 and
the cooling flow component of model~2 are $6.4 \times 10^{42}$
and $2.8 \times 10^{43}$ ergs s$^{-1}$, respectively.
The fraction of the hot component is thus far larger than 
could be explained in terms of the projected 
foreground/background emission from the off-center regions.

\section{SUMMARY AND DISCUSSION}
The energy-dependent flux-mixing effect due to the XRT PSF
makes it quite difficult to analyze the data of extended sources 
observed with {\it ASCA}.
In order to perform simultaneous fitting of the X-ray images 
and spectra,
we have calibrated the XRT PSF using the data from Cyg~X-1,
and have fully taken it into account in the data analysis.

The X-ray radial brightness profile obtained by the {\it ROSAT} PSPC
exhibits a central excess above a single $\beta$-model
within $\sim$$1'.5$, as does the radial profile from 
the {\it Einstein} IPC data.
With {\it ASCA}, we observed the X-ray surface brightness 
in higher energy bands up to $\sim$10~keV for the first time.
Using the newly developed analysis technique, 
we fit the radial brightness profiles and found that 
there is no clear difference among the profiles 
in different energy bands.
This suggests that the central excess emission found 
in the soft X-ray images
from {\it Einstein} and {\it ROSAT} also exists in the higher 
energy band of 4--10~keV.
In particular, we successfully reproduced the GIS radial brightness 
profile as a sum of two $\beta$-models.
The narrower of the two $\beta$ components, thought to represent the
central excess emission, is required not only below 2~keV but also
in harder energy bands, up to 10~keV, assuming a double $\beta$
profile in all energy bands.

The simultaneous fitting of the five annular spectra taken from 
the GIS data
gave the radial profiles of the temperature and metal abundance
(see Fig.~\ref{gis-doujifit-fig}).
The obtained overall temperature structure is consistent with being 
isothermal;
this result is also consistent with the fact that all the radial
profiles are very similar.
However, the PSPC spectra accumulated from the region within $1'.5$,
where the surface brightness profile begin to deviate from the 
$\beta$-model,
gives a significantly lower temperature than that obtained from 
the GIS spectrum. 
This means that there must be an additional cool component 
at the cluster center.
Therefore we jointly analyzed the GIS and PSPC data and successfully 
fit both spectra simultaneously with the two temperature model
as well as the cooling flow model (Fig~\ref{2temp-model-fig}).
The cooling flow model gives the mass deposition rate of $60\pm30$
M$_{\odot}$ yr$^{-1}$,
an order of magnitude smaller than
the 600 M$_{\odot}$ yr$^{-1}$ estimated from the {\it Einstein} data
by David et al. (1990).
If we assumed that all the flux coming from the central $1'.5$ region
is originated from cooling flow,
we would derive a value consistent with 600 M$_{\odot}$ yr$^{-1}$,
using the formula of $\dot{M} = 2\mu m_p L_x / 5kT$
and the bolometric luminosity within $1'.5$ of
$L_x \sim 4 \times 10^{44}$ ergs s$^{-1}$.
However, as we showed in \S~6,
the central $1'.5$ region can not be entirely cooled.
More than 90\% of the total emission measure 
consists of the hot non-cooled component with the temperature 
of $\sim$3.5~keV.
Therefore, the central excess in the X-ray brightness profile
can not be formed only by the cool component.

Since the central region representing the excess emission 
mostly occupied by the same hot ICM component 
which permeates the rest of the cluster,
we interpret the central excess emission as an evidence for
gravitational potential structure.
The potential structure can be interpreted as consisting of two 
distinct components; 
a large scale cluster component and 
a central compact component attributed to the cD galaxy.
In previous investigations, such a dual potential structure has been 
suggested (\cite{TFN87}; \cite{NB95})
or assumed (e.g. \cite{Stewart}).
The first direct observational evidence of the additional potential 
dimple around NGC~1399 in the Fornax cluster was found from 
the {\it ASCA} observation (\cite{Ikebe96}). 
In the case of the Hydra-A cluster, 
there also must be a central potential dimple around the cD galaxy
which primarily causes the central excess in the brightness profile.

Thus far, the central excess brightness seen in many clusters has been
interpreted mainly as due to the central temperature decrease of gas,
thus providing a basis for the cooling flow.
However, our results clearly reveal that 
the central excess brightness is at least
partially caused by the dual potential structure around the cD galaxy.
Therefore, the cooling flow rate derived from the central excess
brightness can be grossly over-estimated, as in the case of 
the Hydra-A cluster, if the dual potential structure is ignored.

The metal abundance distribution is also a very important subject 
which is strongly related to the cluster evolution scenario.
A detailed measurement has become possible for the first time
using {\it ASCA}.
In the Hydra-A cluster, we found an indication of a central 
concentration in the metal abundance distribution 
(Fig.~\ref{gis-doujifit-fig}).
{\it ASCA} has clearly detected the central concentration in
the heavy element abundance distributions in the nearest 
two clusters, 
Virgo and Centaurus (\cite{Matsumoto}; \cite{Fukazawa94}),
which are the firm confirmation of the previous results by {\it Ginga}
(\cite{Koyama90}) and {\it ROSAT} (\cite{AF94}), respectively.
A similar feature has also been discovered in the poor cluster AWM7
(\cite{Xu}).
The central abundance concentration may be caused by a large
contribution by metal-enriched ISM of the cD galaxies, 
which may have not suffered a ram-pressure-stripping process
in the cluster evolution (\cite{Tamura96}), 
because the cD galaxy is sitting in the bottom of the gravitational 
potential well.
We speculate that all the clusters showing central excess emission
in the brightness profile have central concentration 
of heavy elements.

Based on the results from the simultaneous fitting 
of the GIS and PSPC 
spectra with the two temperature model (model 1 in \S~6.2),
we calculated the mass of the cool component gas.
We assumed that the cool component coexists with the hot component
within the $1'.5$ region,
and the local pressure balance is achieved between the cool component
and the hot component as $n_{cool} T_{cool} = n_{hot} T_{hot}$.
The mass of the cool component gas is estimated to be 
$M_{gas} \sim 6 \times 10^9$ M$_{\odot}$; and its filling factor,
that is, the volume fraction of the cool component, 
is $\sim$6 $\times 10^{-4}$.
On the other hand, if the cool component is concentrated 
at the center, and the pressure equilibrium is achieved 
at the boundary of the cool component and the hot component,
the cool component would be distributed out to $\sim$5~kpc 
and would have a total mass of $M_{gas} \sim 2 \times 10^9$ 
M$_{\odot}$.
These values are about 0.1--0.4\% of the stellar mass 
of the cD galaxy 3C218 = Hydra-A, 
estimated to be $1.5 \times 10^{12}M_{\odot}$ 
based on the assumption of $M/L_B=6(M/L_B)_{\odot}$ 
and log$L_B = 11.4$.
This is comparable to that in other non-cD ellipticals (\cite{FJT}).

\acknowledgments
We appreciate the ASCA\_ANL and Sim\_ASCA software development teams for 
supporting to build the data analysis tools.
We also thank all members of the {\it ASCA} team for spacecraft 
operation and data acquisition.
We are grateful to Maxim Markevitch for important comments and
helping PSF calibration, 
and to Joel Bregman for referee's comments.
Y.I. acknowledges support from the Special Researchers' Basic
Science Program and thanks Makoto Hattori for helpful discussion.
M.H. acknowledges support from the Research Fellowships of the Japan
Society for the Promotion of Science for Young Scientists.

\appendix
\section{PSF CALIBRATION}
For the analysis of the extended sources, 
adequate knowledge of the XRT PSF is required.
However, neither the calibration experiment performed on the ground
nor the ray-tracing code could reproduce the PSFs with sufficient 
accuracy, and in-orbit calibration was required.
As we described in \S~4, the PSF has largely extended outskirts and
depends on the X-ray energies.
In addition to these characteristics, 
we should note the difficulties of the PSF modeling 
caused by its lack of cylindrical symmetry and strong dependency 
on the position on the focal plane.
Accordingly we have developed the method to reproduce the PSFs
by interpolating among a set of data from Cyg~X-1
taken at various positions on the focal plane
(\cite{Takahashi95}; \cite{IkebeD}).
Cyg~X-1 was selected because it is one of the brightest point X-ray
sources with a hard spectrum.
With this method we can only reproduce the XRT+GIS PSFs.
Because any point source bright enough to yield a sufficient number 
of signal photons within realistic observing times would cause 
considerable event pile-up in the SIS data,
the XRT+SIS PSFs is difficult to obtain.

The observations of Cyg~X-1 used to produce the XRT+GIS PSFs
were performed in November 1993, November 1994 and May 1995
at several offset positions.
Due to the $5'$ relative misalignment between the two XRTs equipped 
the two GIS sensors, GIS-S2 and GIS-S3, we obtained 14 sample
positions on the focal plane (Fig~\ref{cyg-obs}).
Assuming that the eight quadrants of the two XRTs are identical 
and perfectly symmetric, 
we can have practically 92 PSF data sets.
We have divided each pointing data set into 8 energy bands,
0.5--1~keV, 1--2~keV, 2--3~keV, 3--4~keV, 4--5~keV,
5--6~keV, 6--8~keV and 8--12~keV,
yielding a PSF image data base composed of $8 \times 92$ PSF images.
When we produce a PSF at a given position and energy,
we select the nearest 2--8 sample positions in Fig.~\ref{cyg-obs}
and the nearest 2 energy bands, and linearly interpolate them
in the position and energy space.

We examined the accuracy of the generated PSFs
by comparing the synthesized PSF with the observed data of 3C273,
in terms of their radial profiles centered on the peak.
At the energies above 2~keV, the radial profiles agree reasonably.
On the other hand, below 2~keV band, 
the PSFs generated from Cyg~X-1 data have larger HPD by $\sim 0.'2$
and are significantly wider than those of 3C273.
This may be caused by the intrinsic extension 
in the X-ray emission from
Cyg~X-1 due to the X-ray scattering by interstellar dust grains
(\cite{Mitsuda}).
Therefore we improved the PSFs below 2~keV as follows.
For the 1--2~keV band PSF images in the data base,
we combined the outskirts region ($r>6'$) of the 1--2~keV band with
the core ($r<6'$) of the 2--3~keV band images convolved with
a Gaussian of $\sigma = 0'.125$.
As the 0.5--1~keV band images, we used the 2--3~keV band images
which is smoothed with a Gaussian of $\sigma = 0'.625$.

Fig~\ref{new-psf} shows the comparison of the PSFs thus obtained 
with those of the data from 3C273, in terms of their radial 
profiles centered on the peak.
Based on studies using Monte-Carlo simulation,
we introduced a systematic error 5\% for the model predictions
in the analysis in this paper.
The systematic error of the produced PSFs may be caused by 
the inaccuracy of the assumption that the eight quadrants of 
the two XRTs are identical and symmetric.
Futher, additional observation of Cyg~X-1 would reduce the systematic
error in the interpolation procedure on the focal plane.

\clearpage

\figcaption{The X-ray intensity contour map of the Hydra-A cluster 
obtained with the GIS. 
The background subtracted image was convolved with a Gaussian
of $1'$ FWHM. 
The contours are logarithmic where each step corresponds to 
a multiplicative factor of 2.
\label{gis-image}}

\figcaption{The GIS (a) and SIS (b) spectra extracted from the central
$1.5'$ radius region fitted with a thermal bremsstrahlung plus three
emission lines modified by the photo-electric absorption. 
Crosses are the data and the histograms are the model.
For the display, SIS-S1 data and its model are multiplied by 0.1.
\label{central-spec}}

\figcaption{
The X-ray flux contributions from the sky annular regions
to the individual annular spectra of the GIS are illustrated.
We simulated the GIS data of an isothermal symmetric cluster 
having the radial brightness profile shown in Fig.~8.
The temperature, heavy element abundance, and hydrogen column density
are assumed to be 4~keV, 0.3~solar, and $5 \times 10^{20}$ cm$^{-2}$,
respectively.
On the horizontal axis the radius ranges of extracted spectra 
on the detector are plotted.
The vertical axis shows the contribution fraction from different
sky regions, illustrated with different patterns.
\label{mutual-contrib}}

\figcaption{
The results of the annular spectral fits with a thermal bremsstrahlung
plus three emission lines.
Temperature (a) and equivalent width of Fe-K lines (b) 
are illustrated.
\label{brems+gauss_rpro}}

\figcaption{The results of the annular spectral fits with RS-model.
Temperature (a) and abundance (b) are illustrated.
\label{wagiri-fit}}

\figcaption{The background-subtracted radial brightness profiles 
in different energy ranges obtained by the GIS.
\label{gis-rpro}}

\figcaption{The X-ray intensity contour map obtained with the ROSAT 
PSPC.
The image was smoothed with a Gaussian of $1'$ FWHM. The contours are 
logarithmic where each step corresponds to a multiplicative 
factor of 2.
\label{pspc-image}}

\figcaption{The background-subtracted PSPC radial profile fitted 
with the double $\beta$-model. 
Open circles are the data and histograms are the models.
\label{pspc-rpro-fit}}

\figcaption{The best fit normalization ratios between narrow $\beta$ 
and wide $\beta$ components.
The errors represent 90 \% confidence limit.
\label{norm-ratio}}

\figcaption{The radial profiles of the temperature (a) 
and abundance (b)
derived from the simultaneous fitting of the five GIS annular spectra.
\label{gis-doujifit-fig}}

\figcaption{Simultaneous fitting of the PSPC and GIS spectra extracted
from the central $1.5'$ radius region.
Filled circles are the PSPC data and crosses are the GIS data.
Histograms are the best fit two temperature model.
\label{2temp-model-fig}}

\figcaption{The $\chi^2$ contour map (a) for the temperature of the
cool component ($T_{cool}$) vs the fraction of the cool component 
($F_{cool}$, see text) and 
(b) for the mass deposition rate ($\dot{M}$) vs maximum temperature
($T_{max}$).
\label{f-tcool}}

\figcaption{The positions on the focal plane where the Cyg~X-1
is observed (filled circles) and the positions which are identical to
those assuming symmetry (open circles).
\label{cyg-obs}}

\figcaption{Comparison of the radial brightness profiles of the 
3C273 data with those of the PSFs that we used in our analysis.
The PSFs are reproduced by interpolating among 
the actual observed data of Cyg~X-1 or their modified images
(see text).
The vertical axis shows the radial profile ratios (data/PSF)
averaged over eight data sets of 3C237.
\label{new-psf}}

\clearpage

\begin{deluxetable}{cccc}
\tablewidth{0pc}
\tablecaption{The results of the GIS radial profile fitting
with a single $\beta$-model.
\label{rprofit}}
\tablehead{
\colhead{energy range(keV)}  &  \colhead{$r_c$(arcmin)}  &
\colhead{$\beta$}       &  \colhead{$\chi^2/\nu$}}
\startdata
0.5--1.5   & 0.97 (0.71-1.25) & 0.62 (0.58-0.66) & 10.5/17 \nl
1.5--2.5   & 0.96 (0.70-1.30) & 0.63 (0.59-0.68) & 15.1/17 \nl
2.5--4     & 1.14 (0.69-1.35) & 0.67 (0.58-0.68) & 7.0/17  \nl
4--6       & 1.54 (1.10-2.25) & 0.72 (0.64-0.87) & 5.6/17  \nl
6--10      & 1.20 (0.58-2.05) & 0.67 (0.56-0.87) & 7.3/17  \nl
\enddata
\tablecomments{
The errors represent 90\% confidence limit
for two parameters of interest.}
\end{deluxetable}

\begin{deluxetable}{cccccc}
\tablewidth{0pc}
\tablecaption{Simultaneous fitting of the five GIS 
annular spectra
\label{gis-doujifit}}
\tablehead{
\colhead{Radius(arcmin)} & \colhead{0--1.5} & \colhead{1.5--3} &
\colhead{3--5} & \colhead{5--10} & \colhead{10--20}}
\startdata
Temperature(keV)  & 3.41$^{+0.20}_{-0.25}$  & 4.26$^{+1.11}_{-0.81}$  & 
4.08$^{+1.09}_{-1.18}$ & 3.62$^{+0.98}_{-0.60}$  & 2.77$^{+0.75}_{-0.60}$ \nl
Abundance(solar)  & 0.57$^{+0.25}_{-0.21}$ & 0.17 ($<$0.55) &
0.015 ($<$0.74)   & 0.25 ($<$0.71)   & 0.25 ($<$0.94) \nl
$N_{\rm H}$ ($\times 10^{20}$ cm$^{-2}$) &
\multicolumn{5}{c}{1.7 ($<$ 3.9)} \nl
$\chi^2/\nu$      & \multicolumn{5}{c}{237.1/283}       \nl
\enddata
\tablecomments{
The errors represent 90\% confidence 
limit for one parameter of interest.}
\end{deluxetable}

\begin{deluxetable}{ccc}
\tablewidth{0pc}
\tablecaption{Obtained temperatures and abundances 
in the central $1'.5$ region and the outside
\label{gis-doujifit-2}}
\tablehead{
\colhead{Radius(arcmin)} & \colhead{0--1.5} & \colhead{1.5--20}}
\startdata
Temperature(keV)  & 3.55$\pm0.18$          & 3.91$^{+0.24}_{-0.23}$ \nl
Abundance(solar)  & 0.56$^{+0.18}_{-0.16}$ & 0.14$\pm0.11$          \nl
$N_{\rm H}$ ($\times 10^{20}$ cm$^{-2}$) &
\multicolumn{2}{c}{1.2 ($<$ 3.2)} \nl
$\chi^2/\nu$      & \multicolumn{2}{c}{247.3/289} \nl
\enddata
\tablecomments{
The errors represent 90\% confidence limit
for one parameter of interest.}
\end{deluxetable}

\begin{deluxetable}{cccccccc}
\tablewidth{0pc}
\tablecaption{Simultaneous fitting of the PSPC and GIS spectra
with a two temperature model
\label{2temp-model}}
\tablehead{
\colhead{plasma} & \colhead{$T_{cool}$} & \colhead{$T_{hot}$} &
\colhead{EM($T_{cool}$)} & \colhead{EM($T_{hot}$)} & 
\colhead{abundance} & \colhead{$N_h$} & \colhead{$\chi^2/\nu$} \\
\colhead{code}   & \colhead{(keV)}      & \colhead{(keV)}     &
\colhead{(cm$^{-3}$)} & \colhead{(cm$^{-3}$)}   &
\colhead{(solar)}  & \colhead{(cm$^{-2}$)} & \colhead{}}
\startdata
RS    & 0.67 & 3.45 & $3.9 \times 10^{65}$ &
$2.4 \times 10^{67}$ & 0.51 & $3.4 \times 10^{20}$ & 74.15/58 \nl
MEKA  & 0.50 & 3.41 & $4.1 \times 10^{65}$ & 
$2.4 \times 10^{67}$ & 0.56 & $3.4 \times 10^{20}$ & 73.97/58 \nl
\enddata
\tablecomments{Both the PSPC and GIS spectra were extracted from 
the central $1'.5$ region.
The fitting model is expressed with sum of two RS-model or 
MEKA-model modified by photo-electric absorption.}
\end{deluxetable}

\begin{deluxetable}{cccccccc}
\tablewidth{0pc}
\tablecaption{Simultaneous fitting of the PSPC and GIS spectra
with a cooling flow model
\label{cflow}}
\tablehead{
\colhead{plasma}  & \colhead{$\dot{M}$} & \colhead{$T_{hot}$} &
\colhead{$EM(CF)$} & \colhead{$EM(T_{hot})$} & \colhead{abundance} &
\colhead{$N_{\rm H}$} &  \colhead{$\chi^2/\nu$} \\
\colhead{code} & \colhead{(M$_{\odot}$ yr$^{-1}$)} & \colhead{(keV)}&
\colhead{(cm$^{-3}$)} & \colhead{(cm$^{-3}$)} & \colhead{(solar)} &
\colhead{(cm$^{-2}$)} & \colhead{}}
\startdata
RS      & 57 & 3.57 & $2.0 \times 10^{66}$ & $2.3 \times 10^{67}$ &
0.50 & 3.3 $\times 10^{20}$ & 73.85/59 \nl
MEKA    & 62 & 3.56 & $2.2 \times 10^{66}$ & $2.2 \times 10^{67}$ &
0.51 & 3.4 $\times 10^{20}$ & 73.72/59 \nl
\enddata
\end{deluxetable}

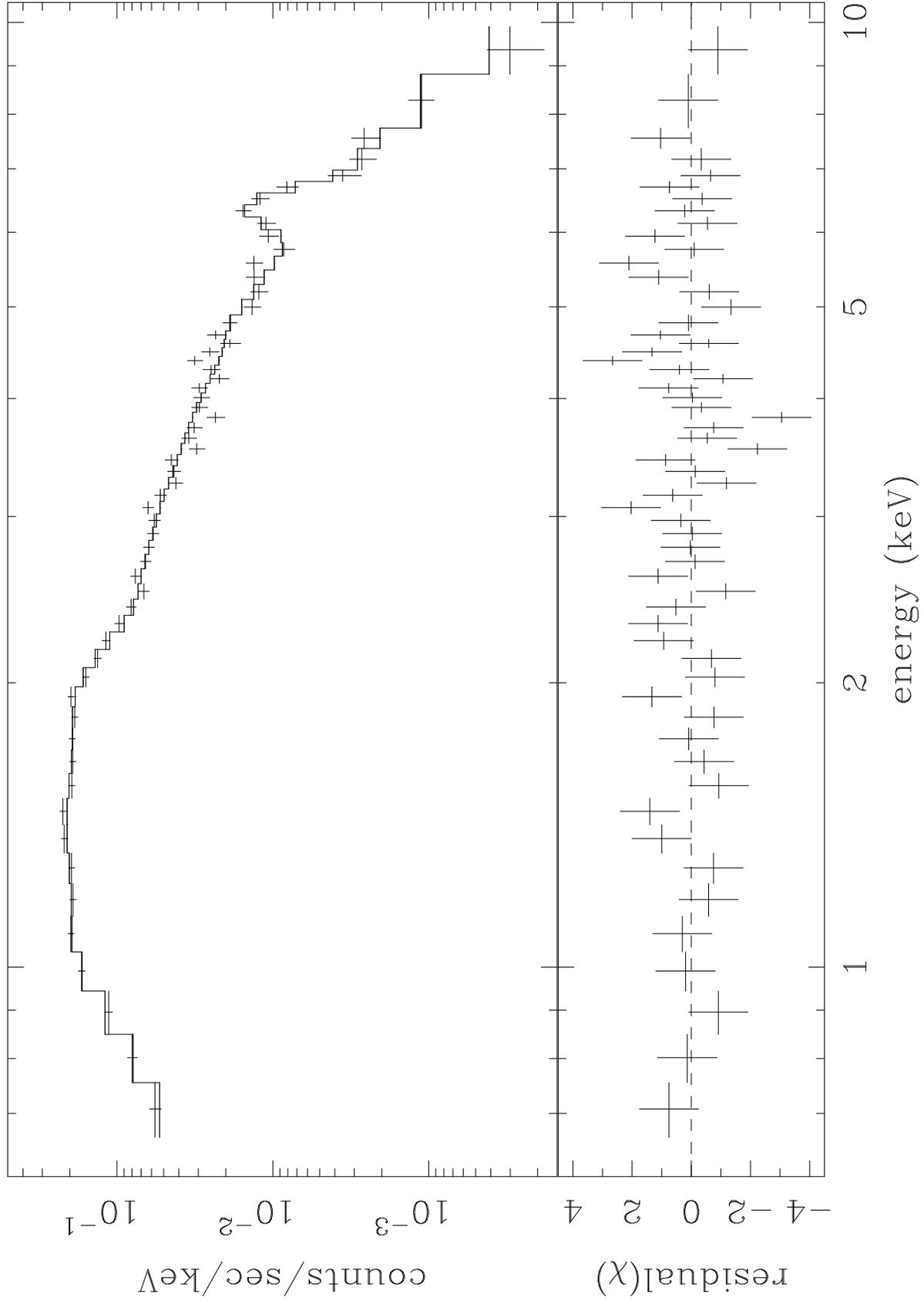
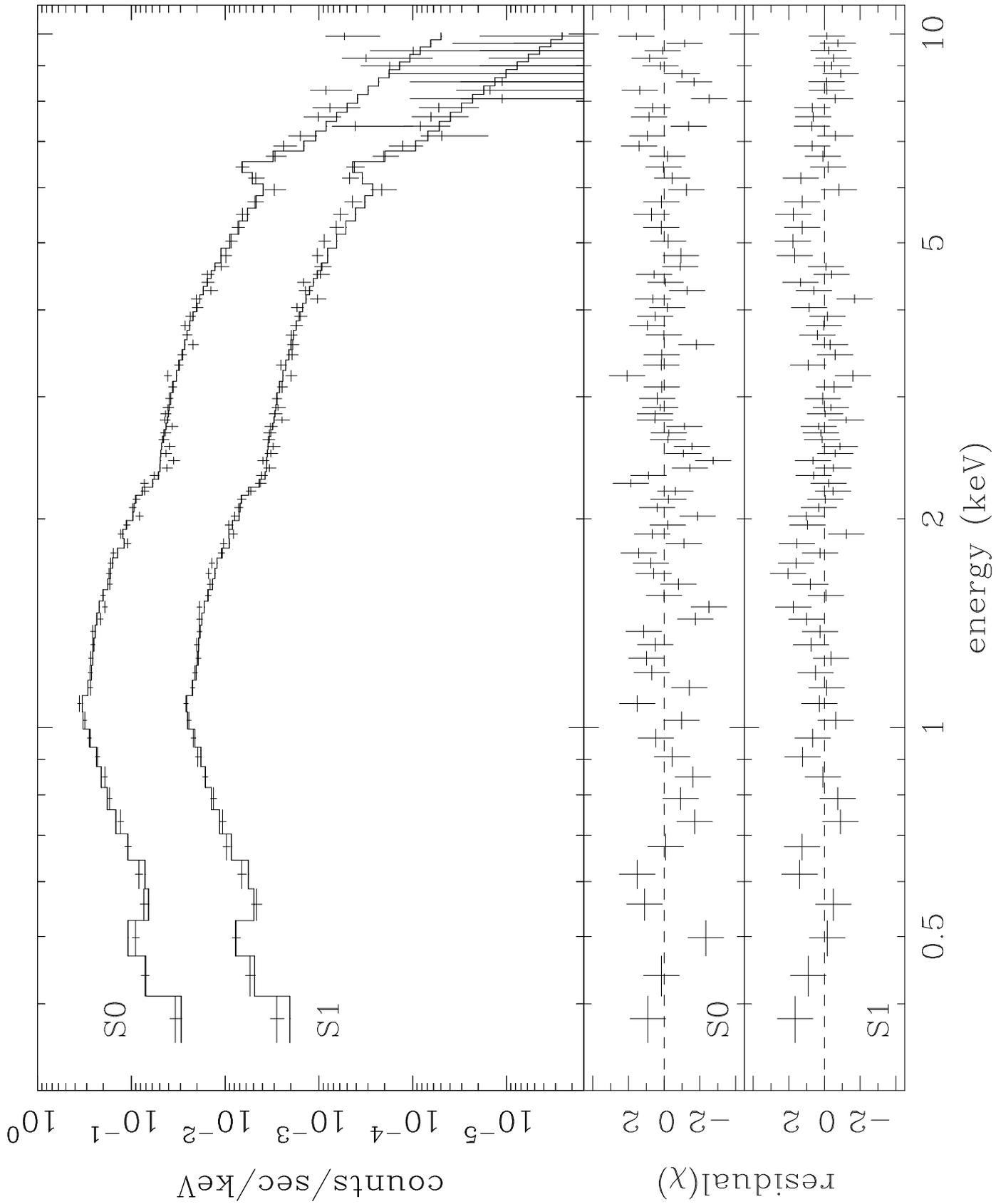
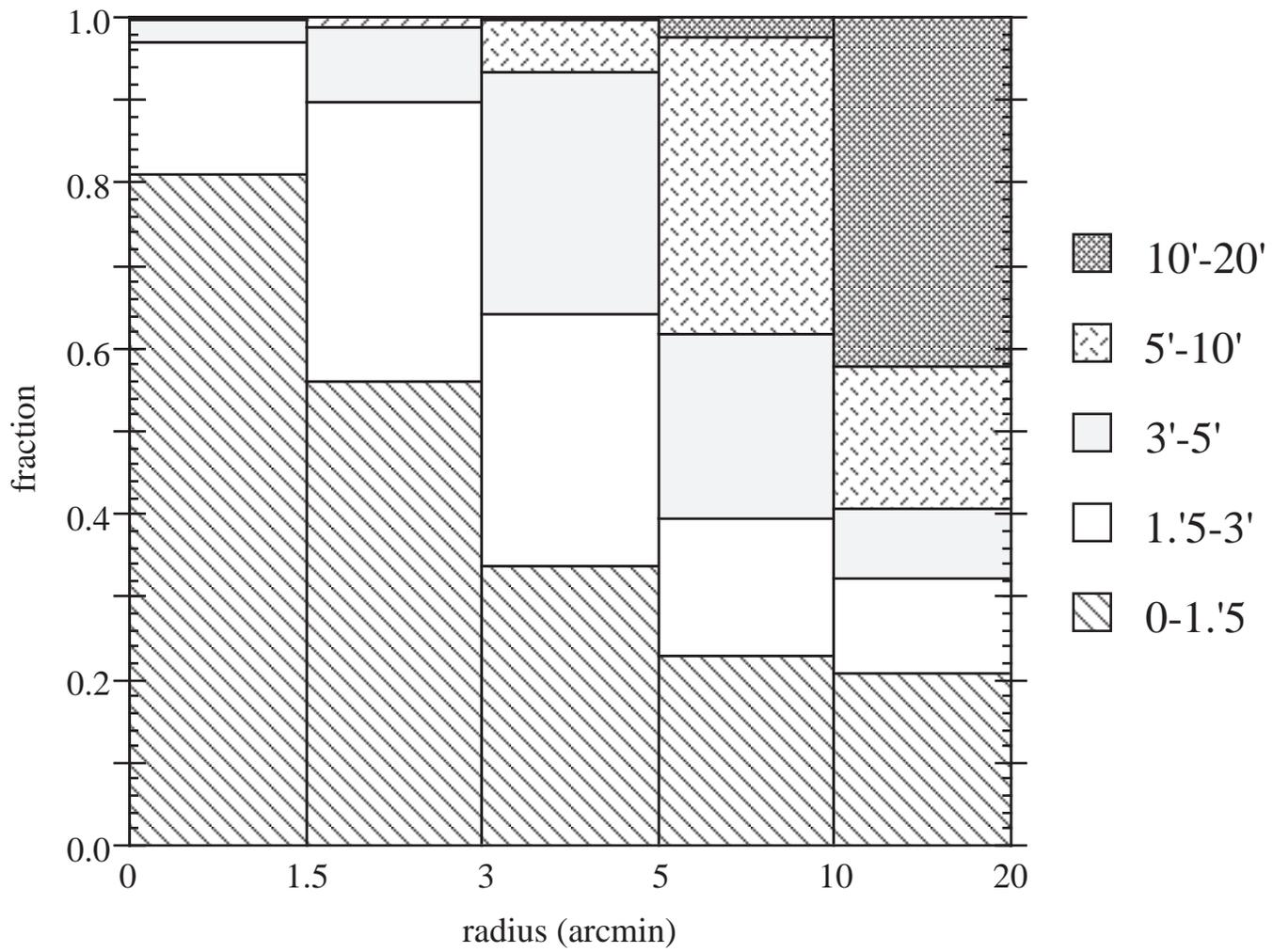
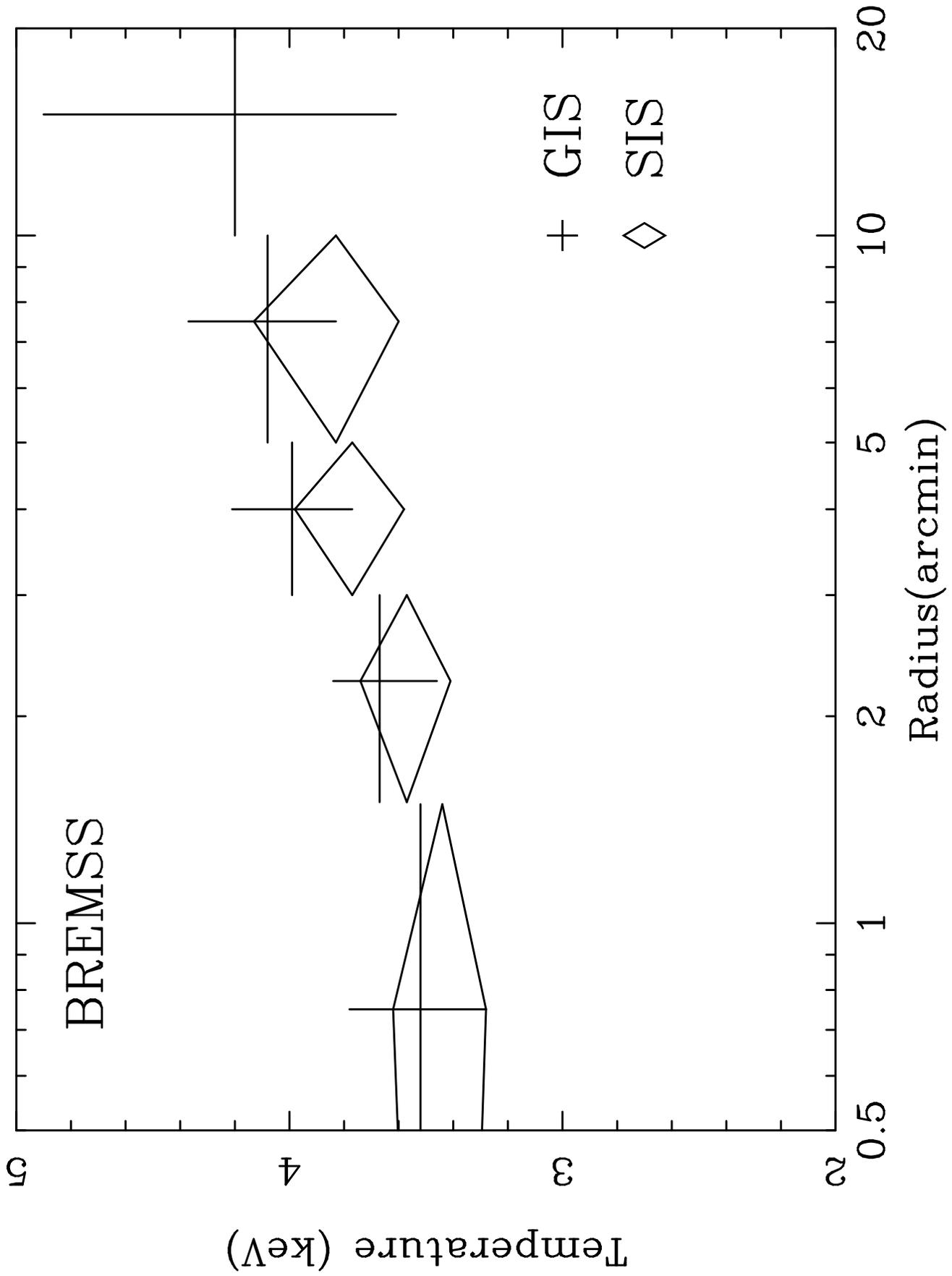
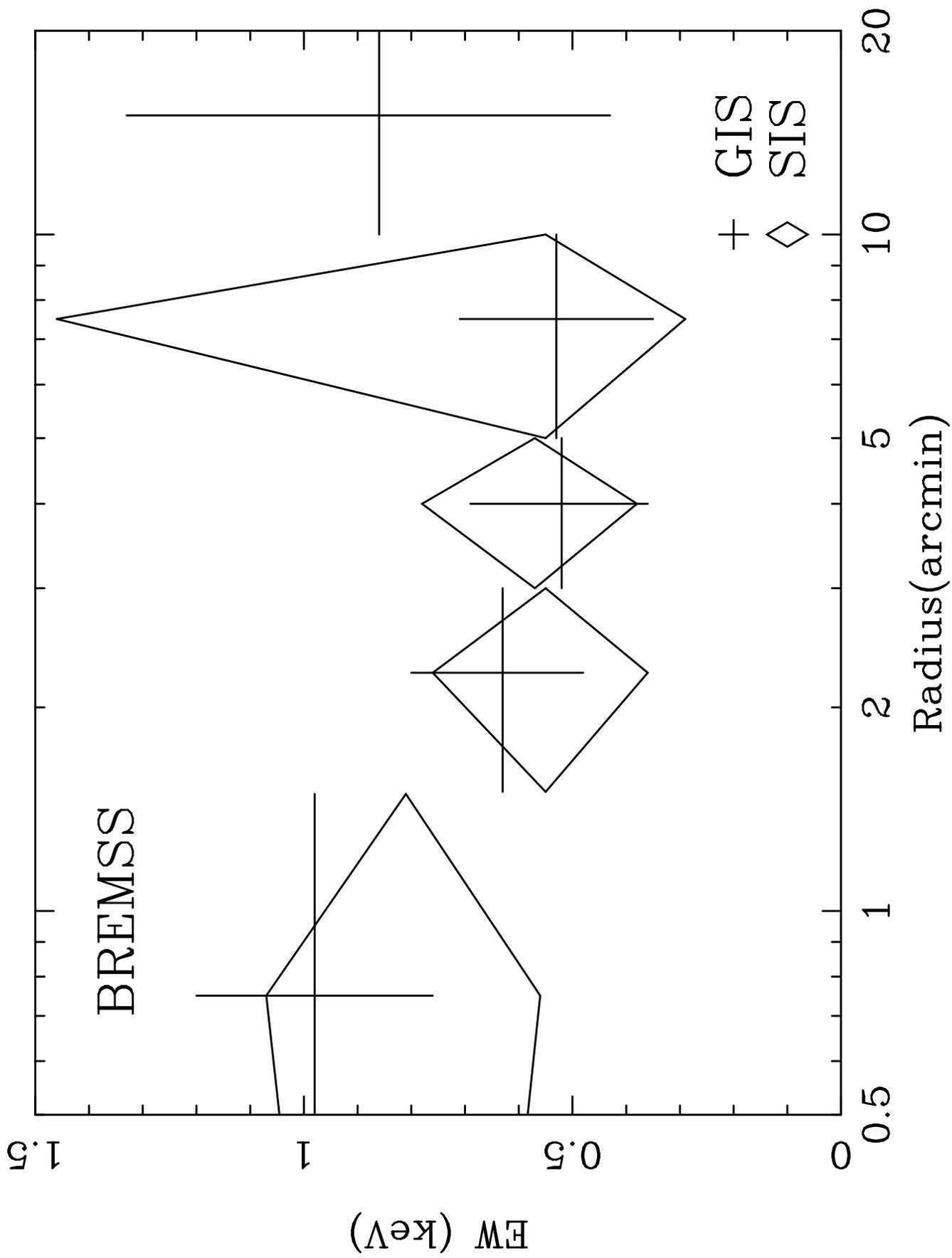
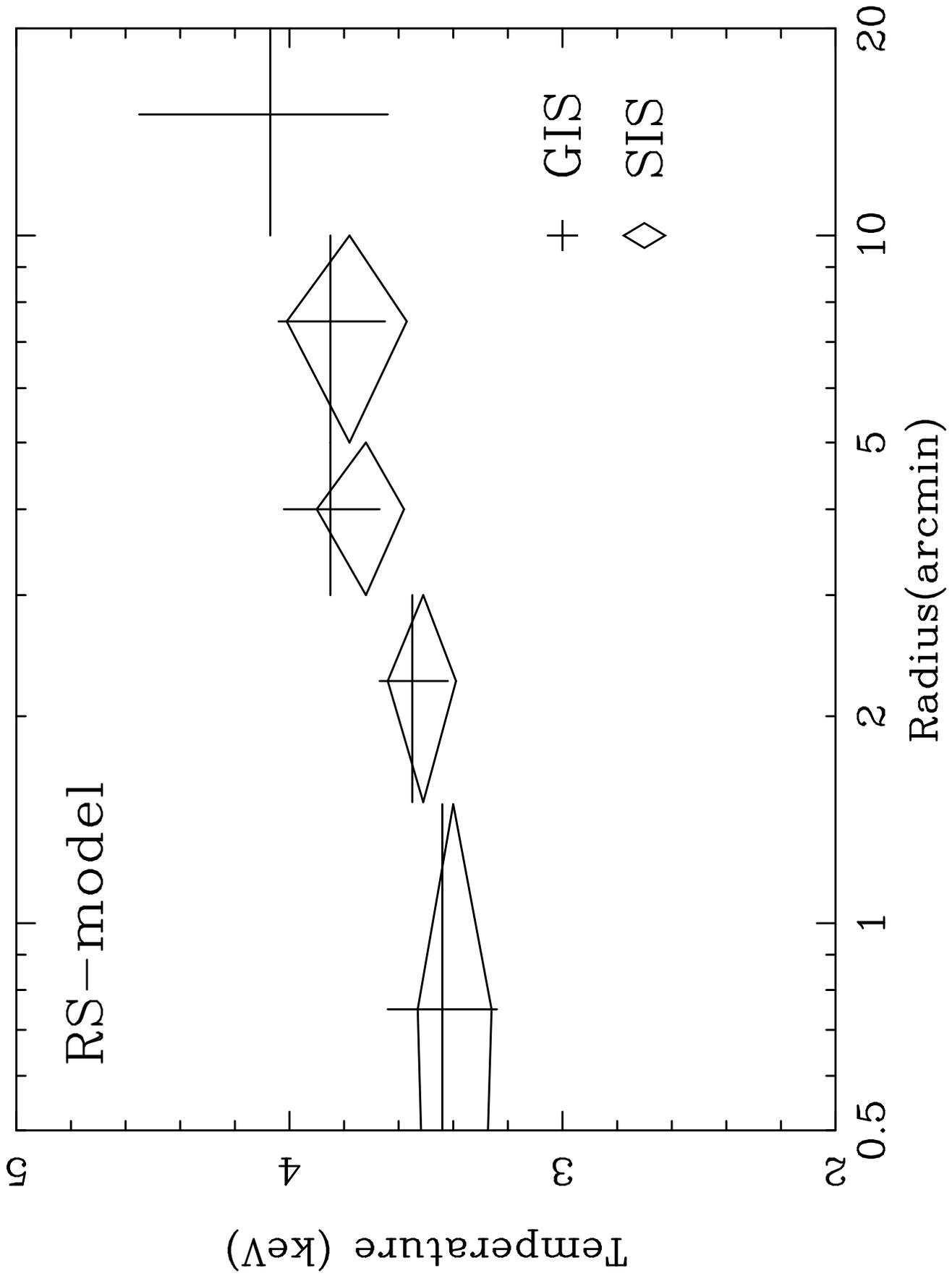
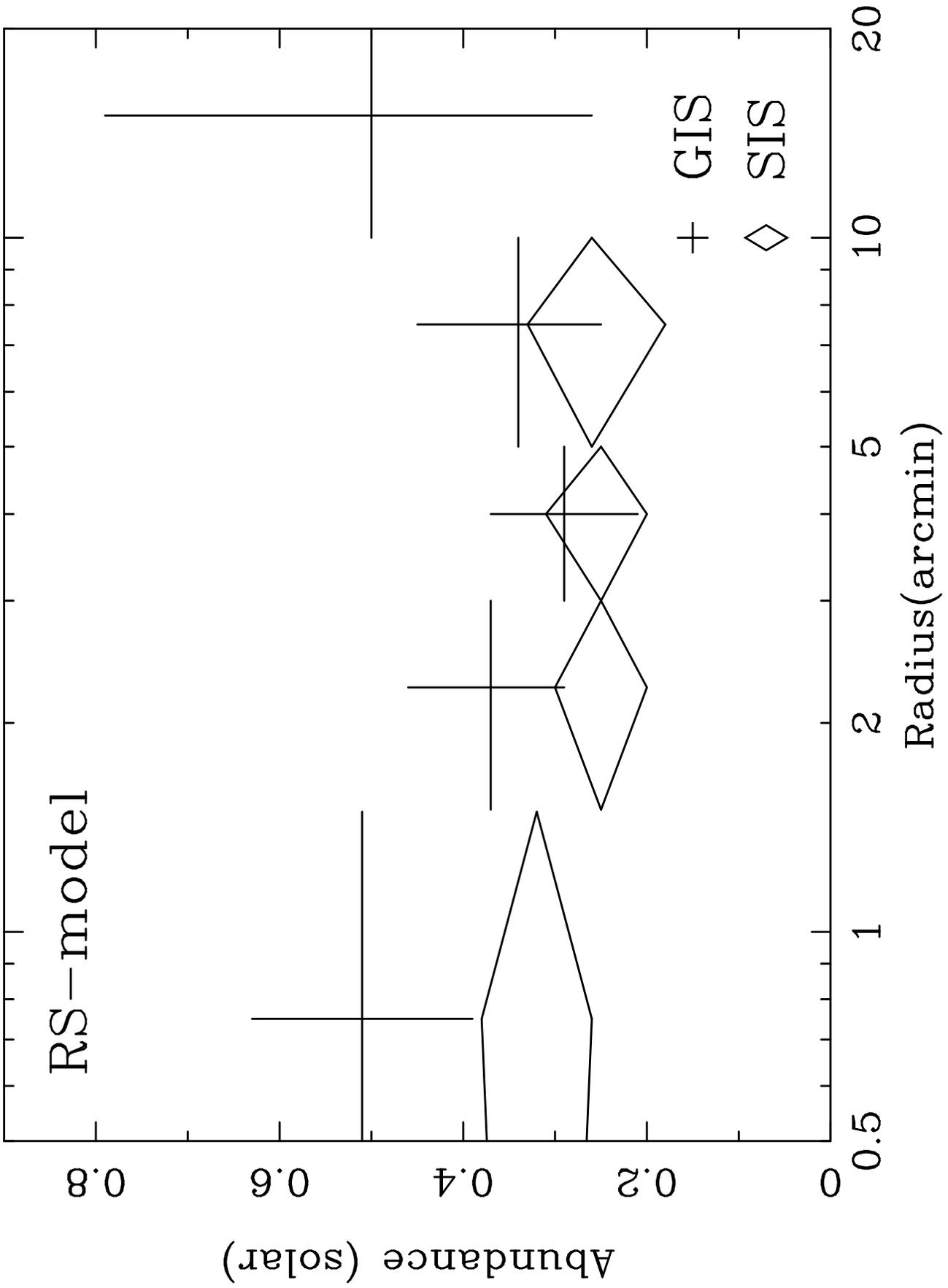
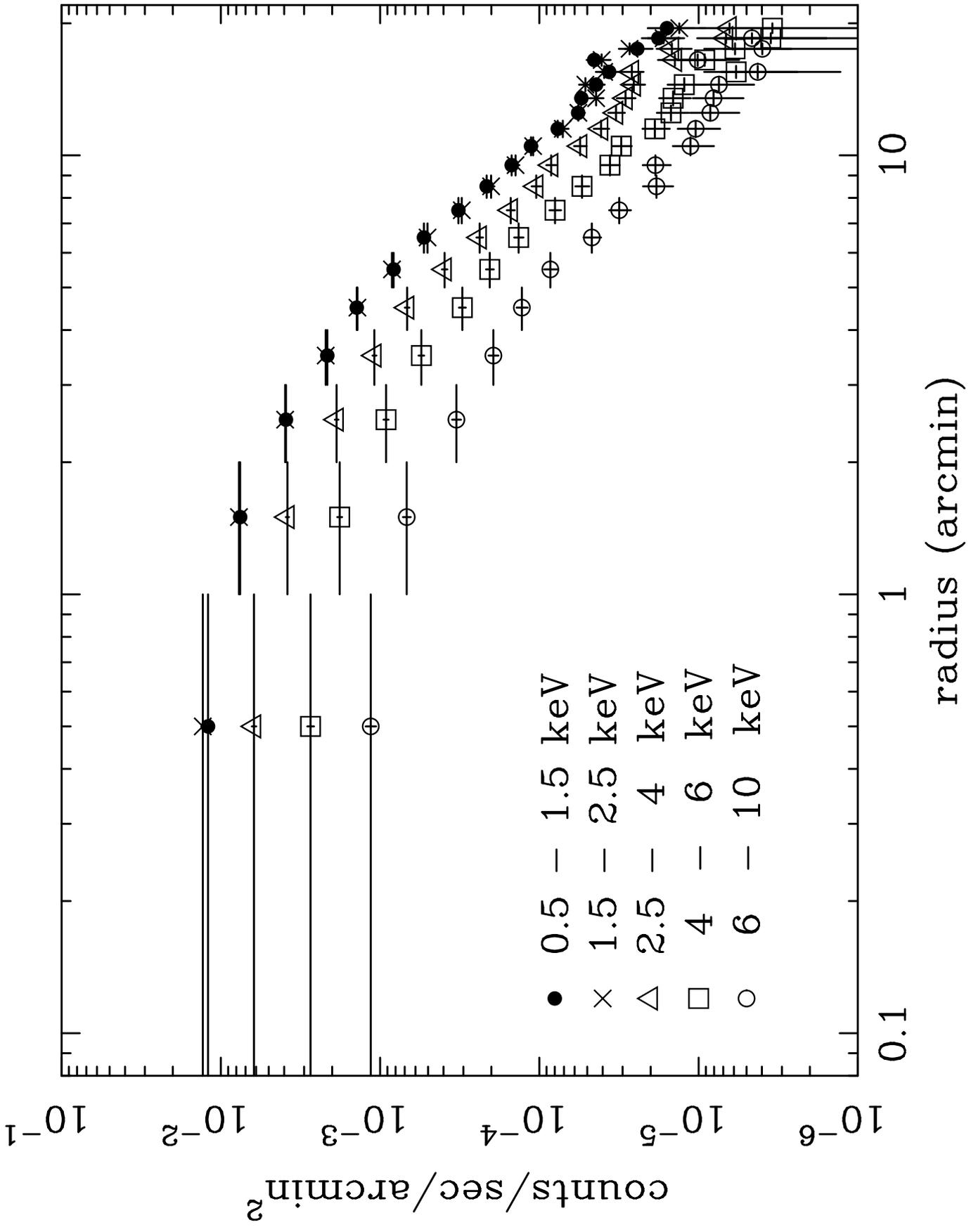
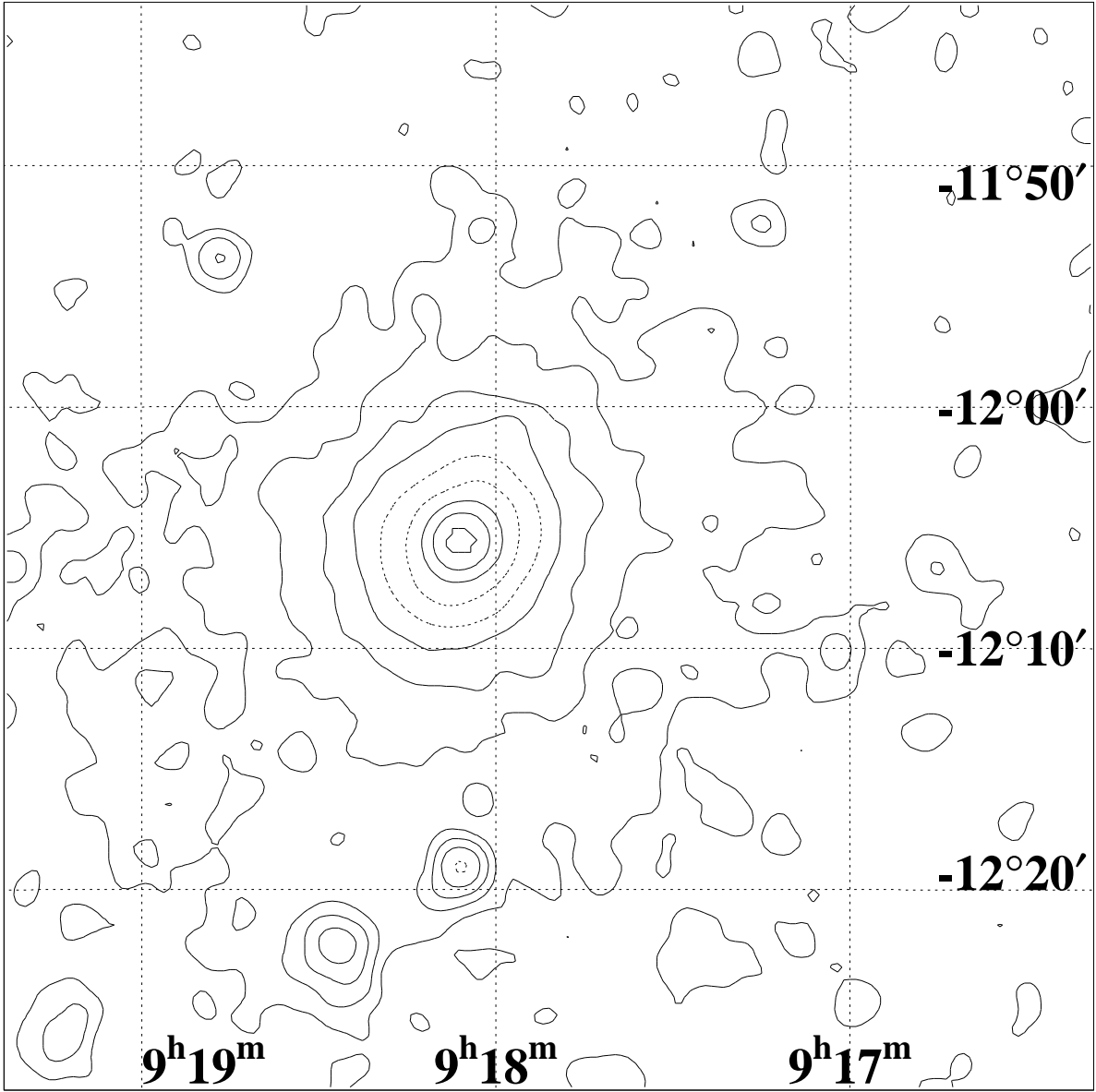
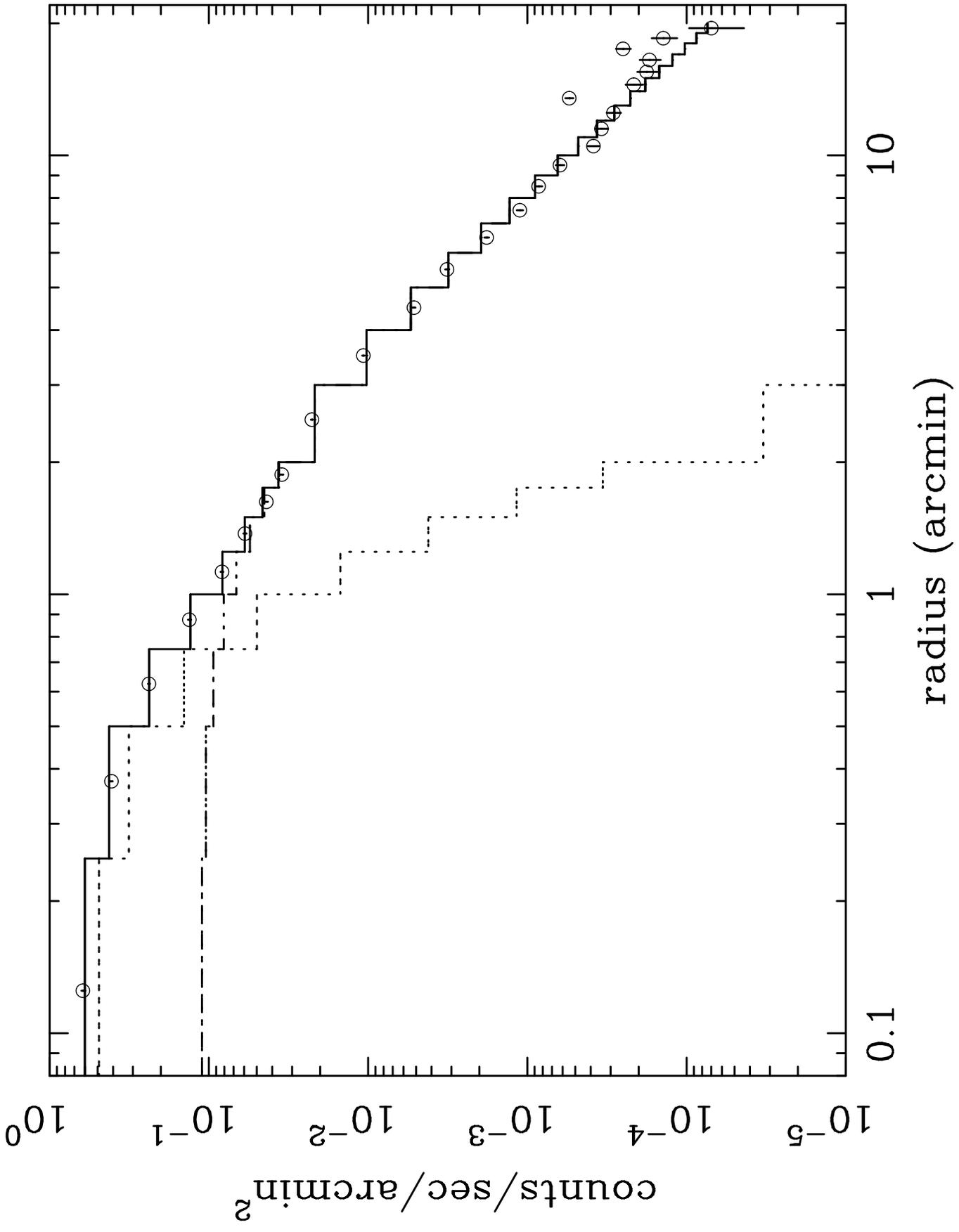
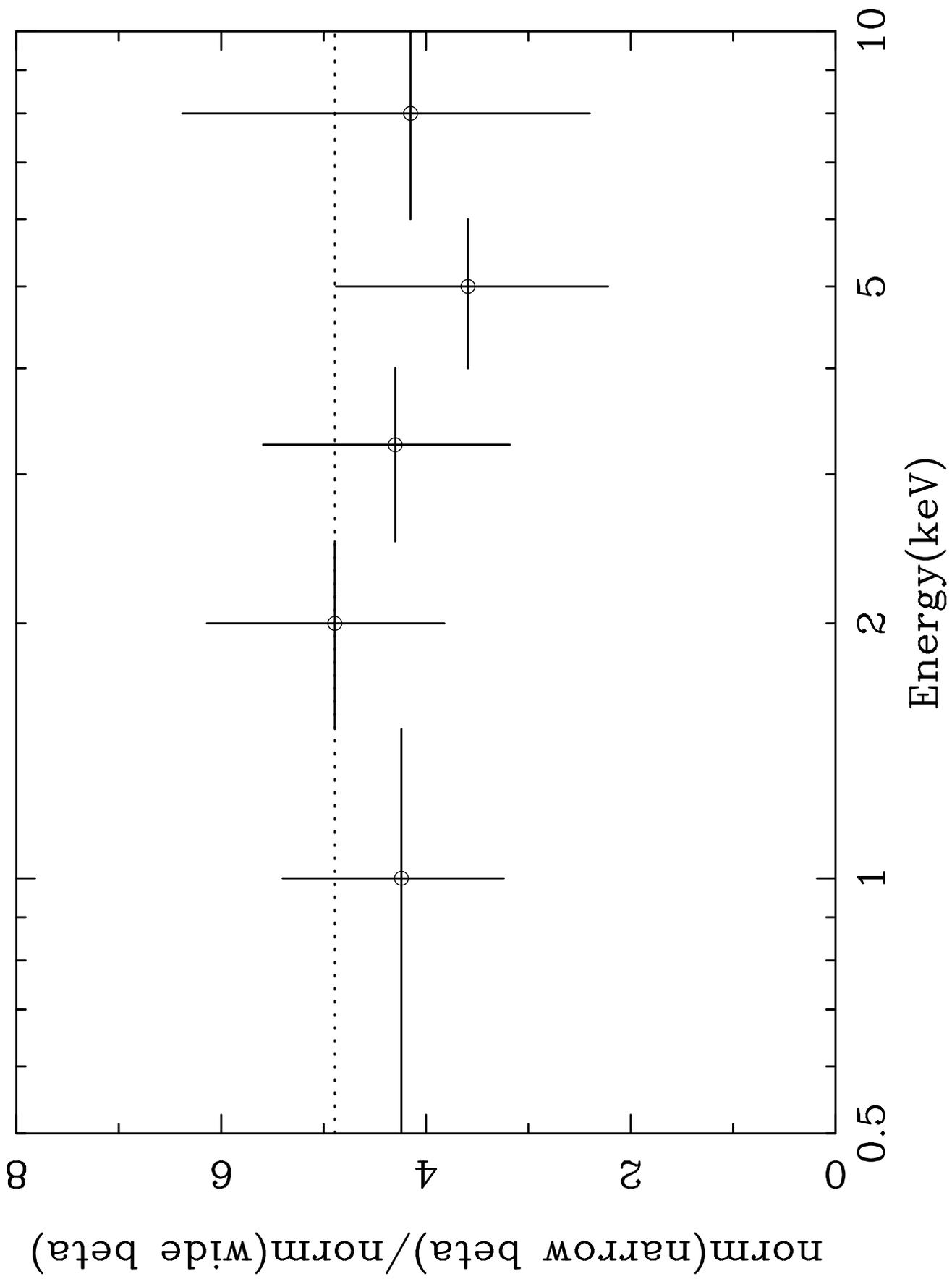
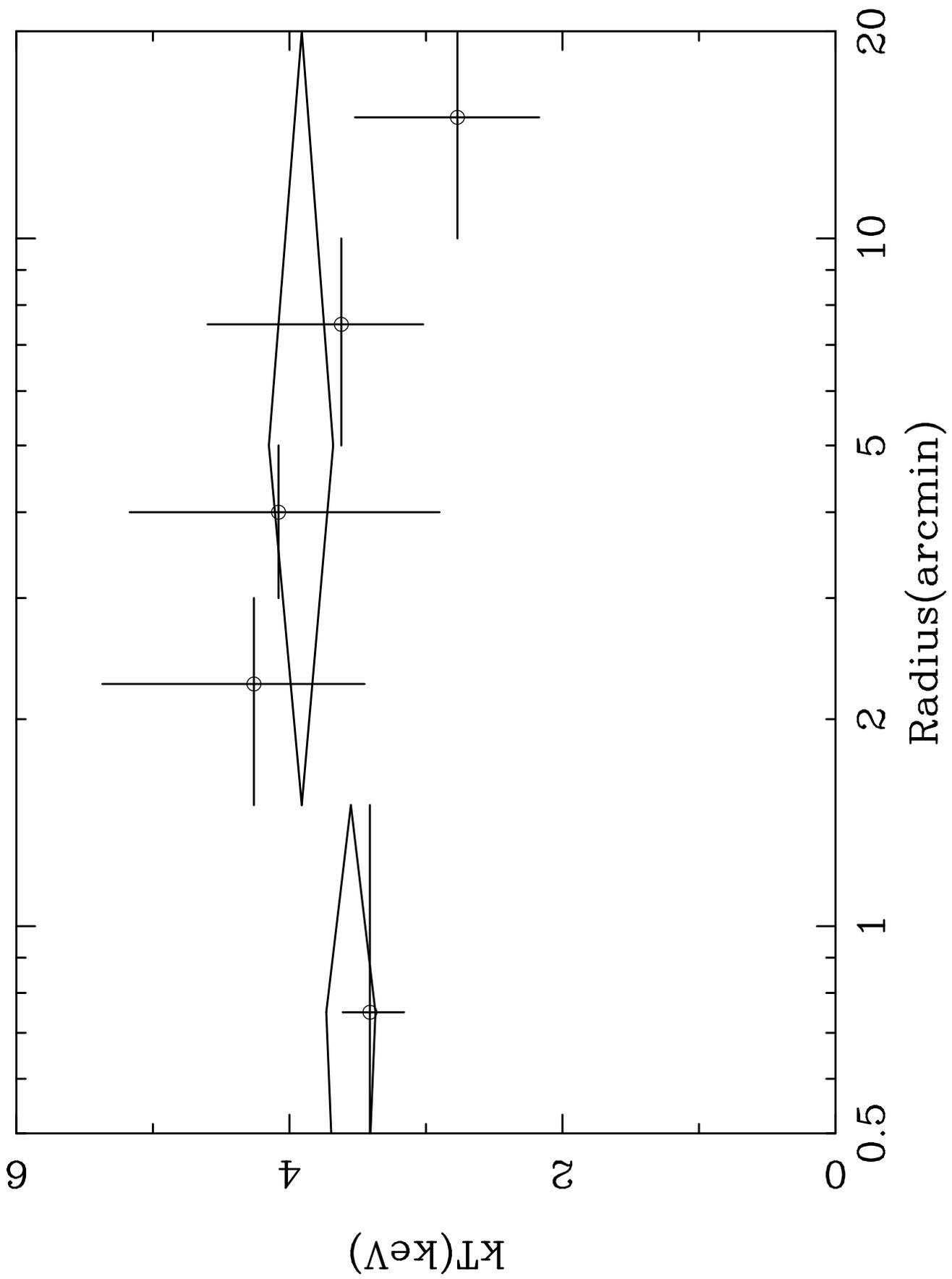
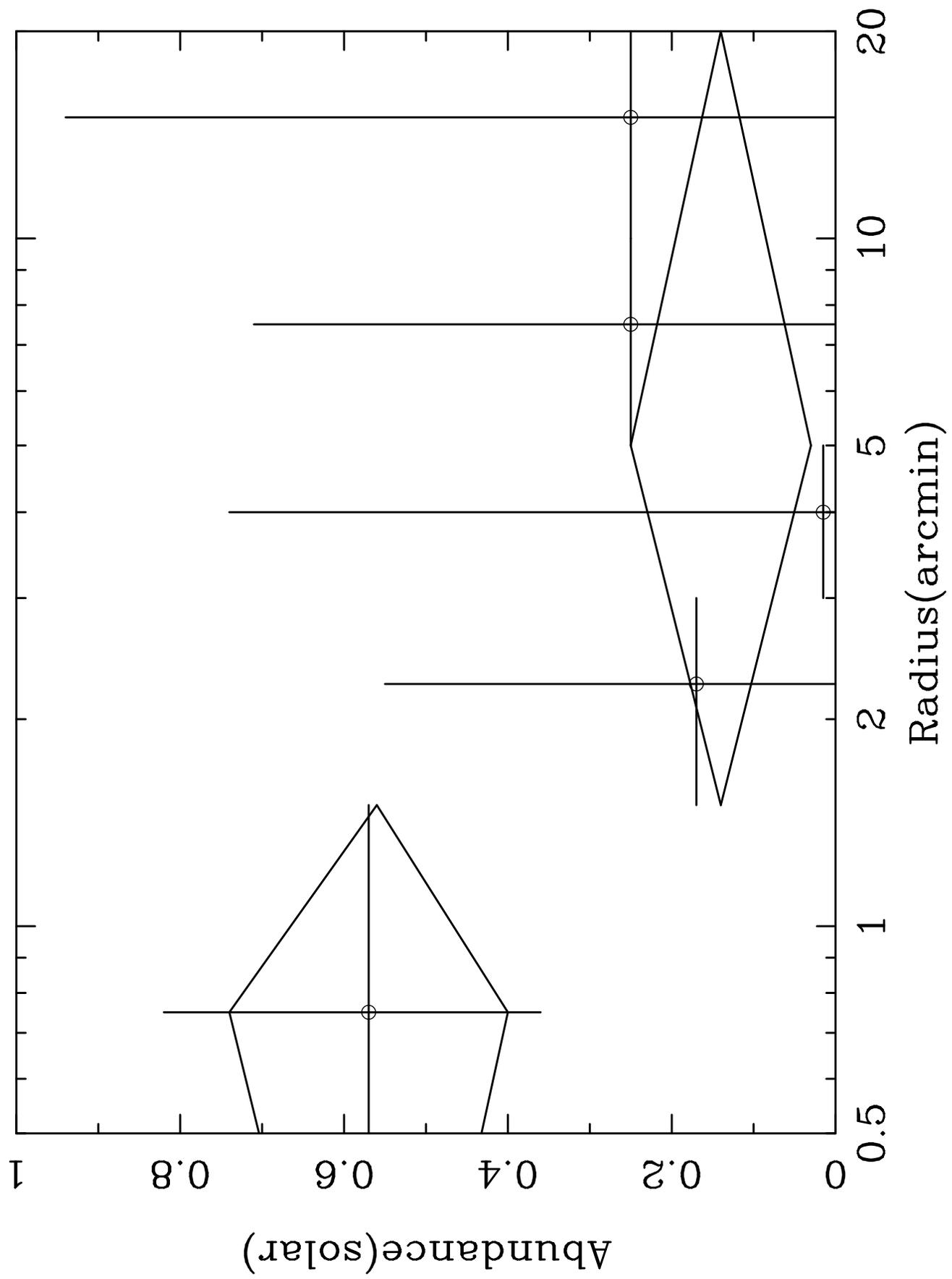
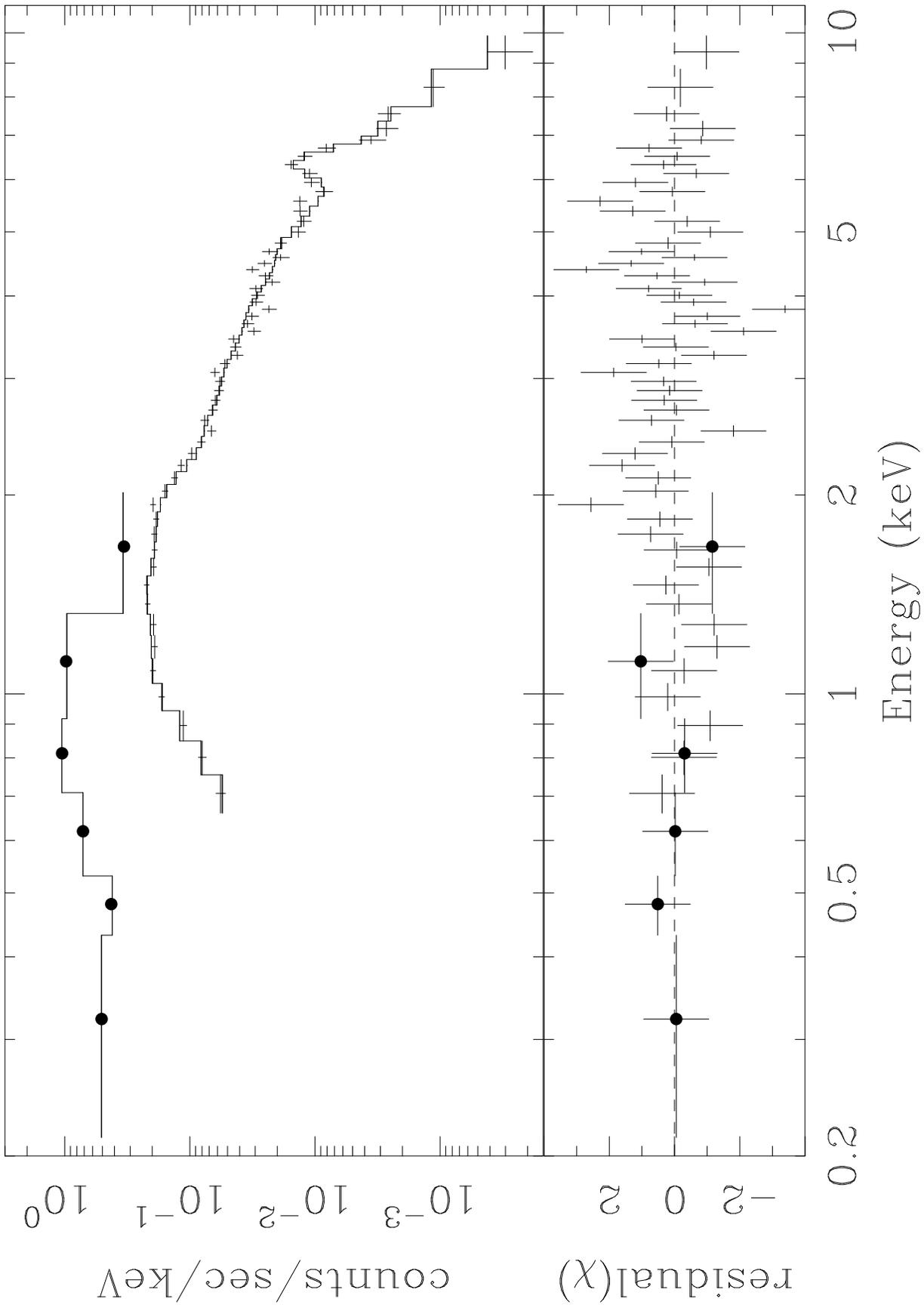
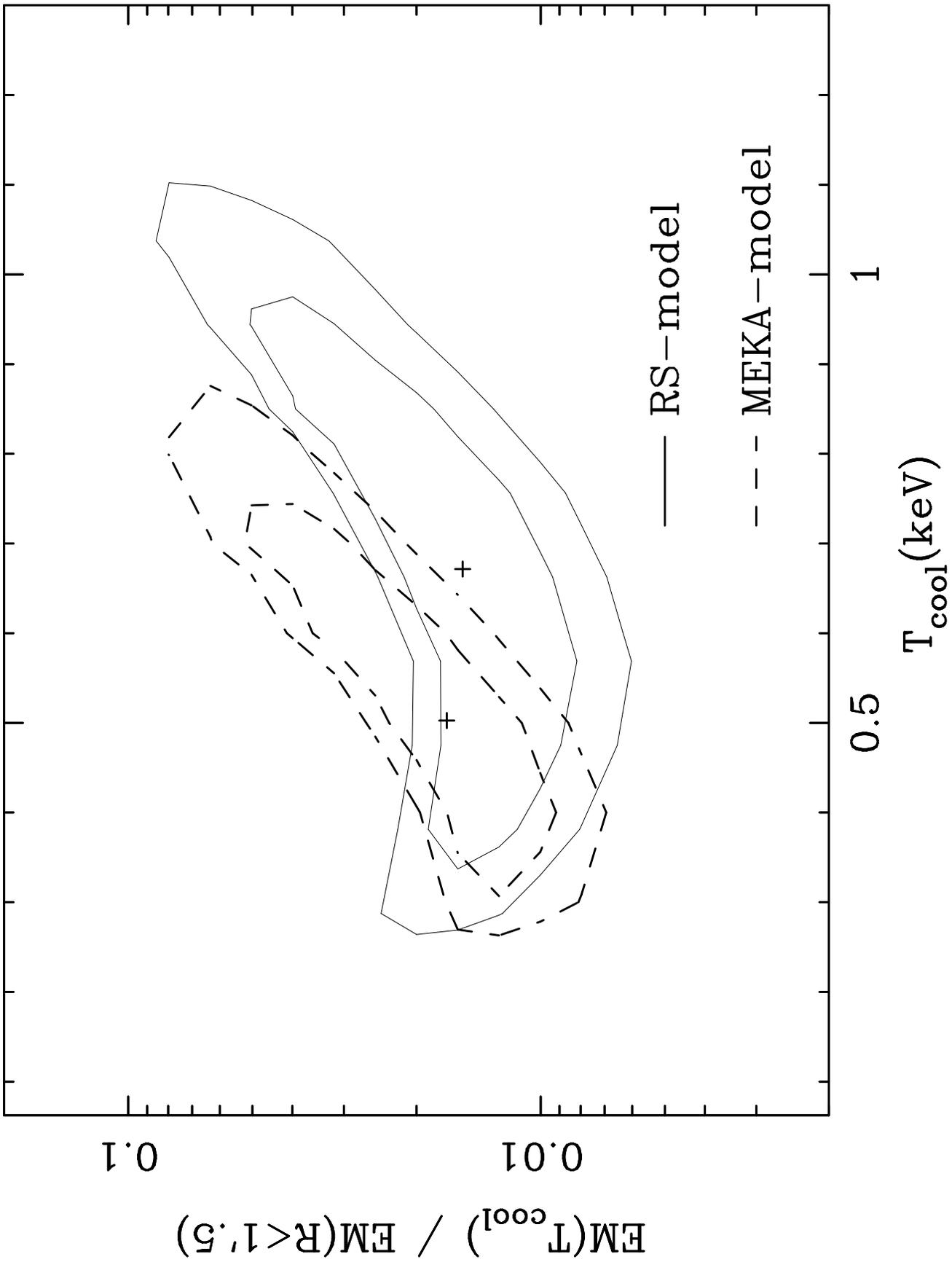
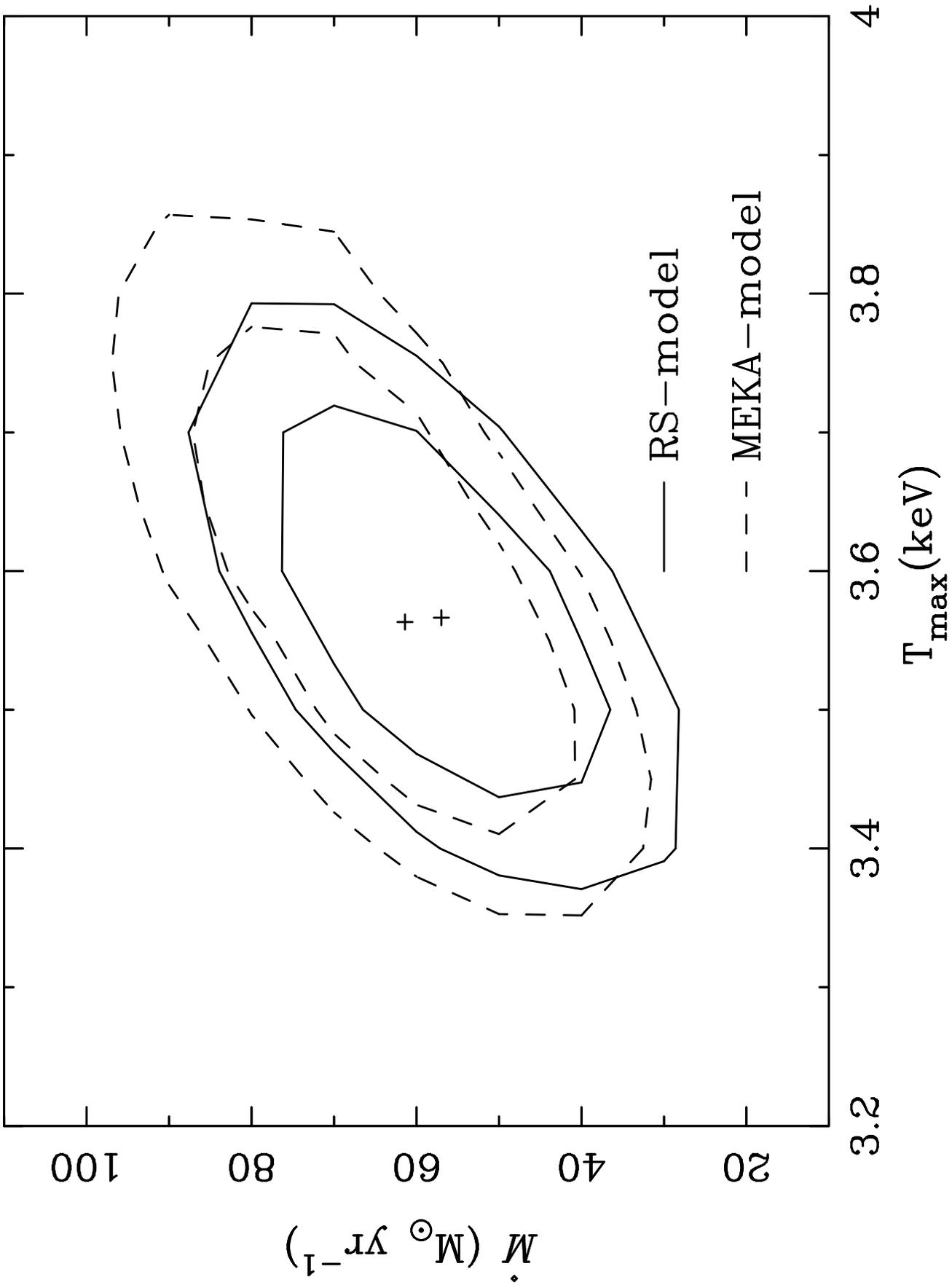
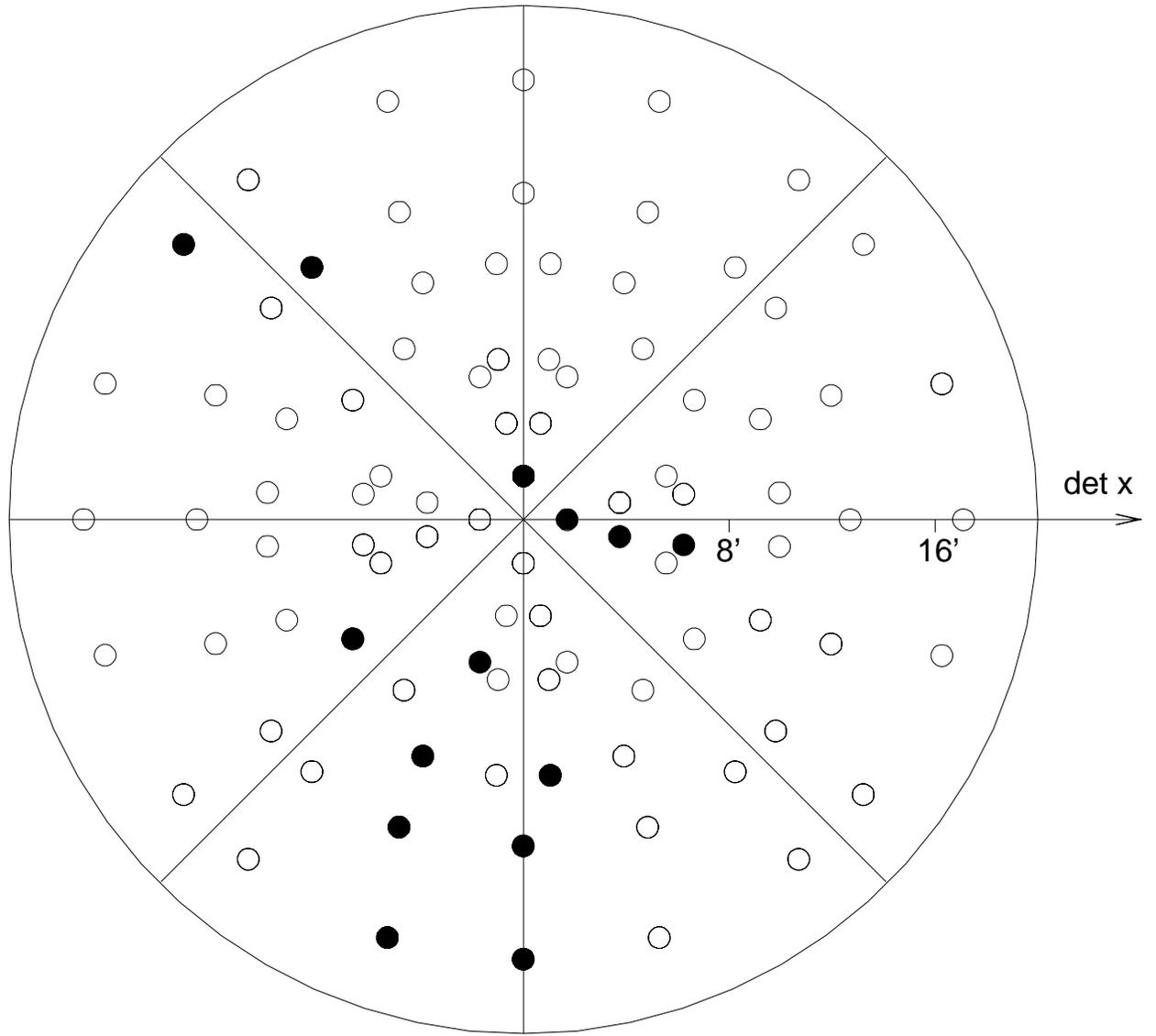
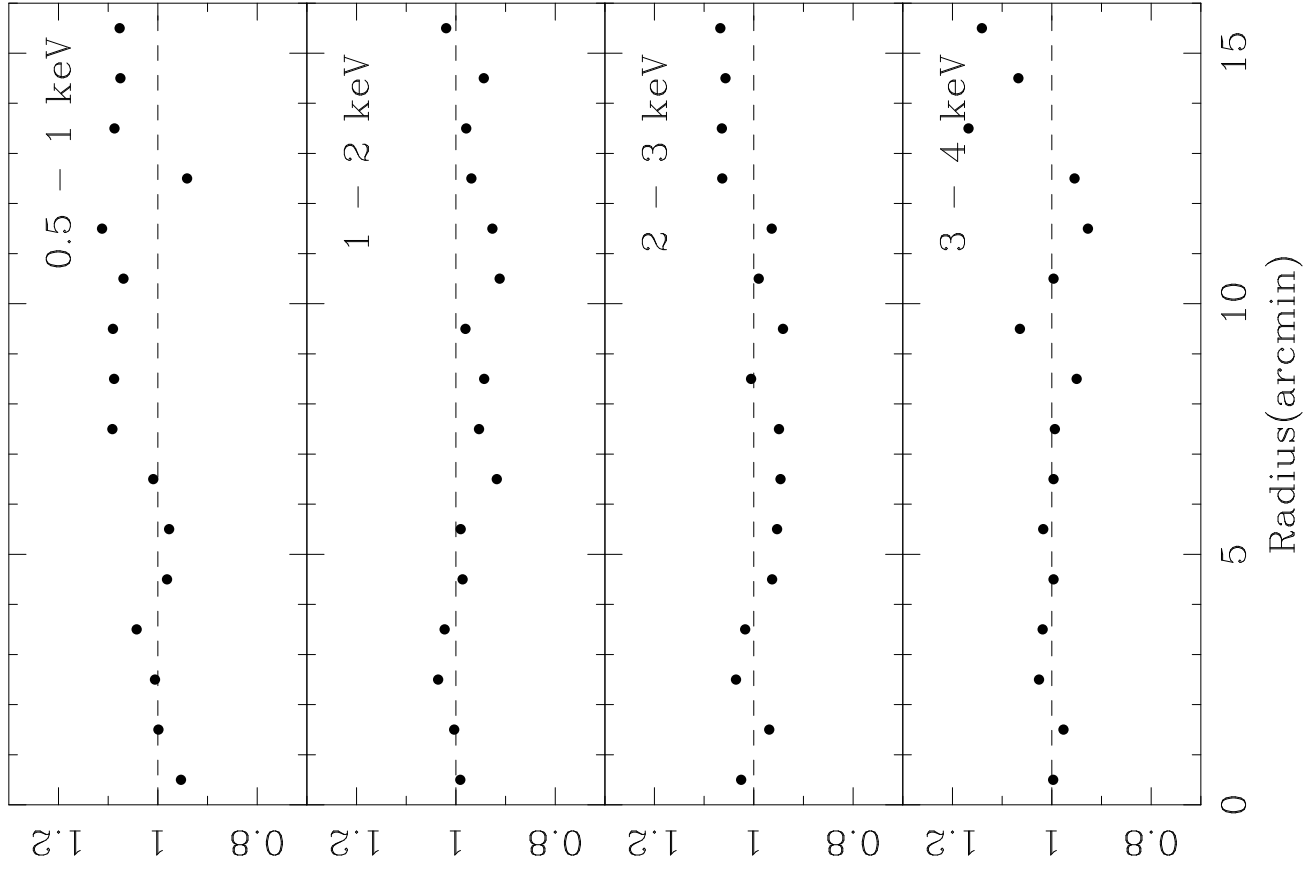
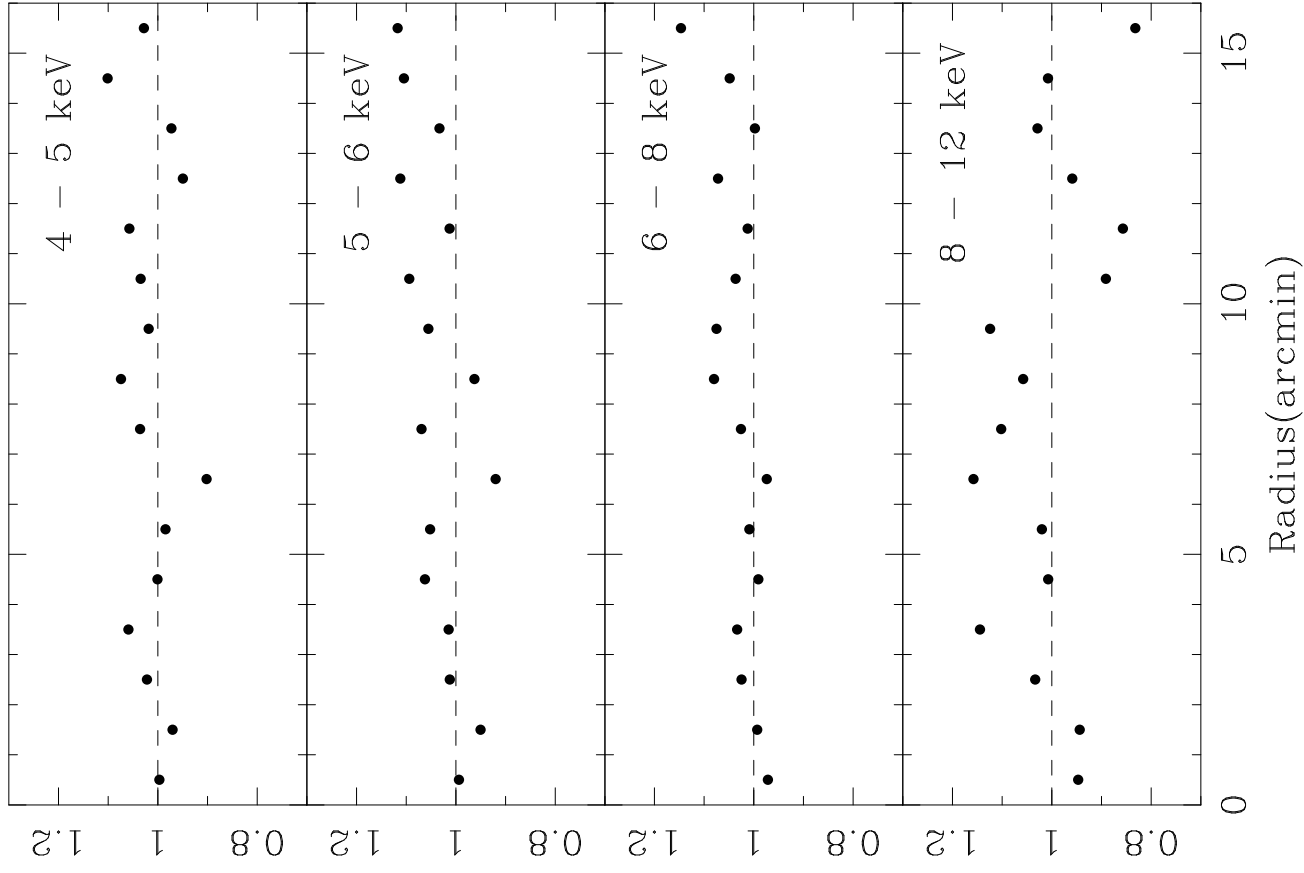
\end{document}